\documentclass[preprint]{aastex}

\newcommand{\apg}{\gtrsim}
\newcommand{\apl}{\lesssim}
\newcommand{\etal}{et al.}
\newcommand{\lya}{\mbox{${\rm Ly}\alpha$}}

\begin{document}

\lefthead{Chen \etal}
\righthead{}


\title{The Las Campanas Infrared Survey. IV. The Photometric Redshift Survey 
and the Rest-frame $R$-band Galaxy Luminosity Function at $0.5 \leq z \leq 1.5$
}

\author{HSIAO-WEN CHEN\altaffilmark{1,2}, 
RONALD. O. MARZKE\altaffilmark{3},
PATRICK. J. McCARTHY\altaffilmark{1}, 
P. MARTINI\altaffilmark{1},
R. G. CARLBERG\altaffilmark{4}, 
S. E. PERSSON\altaffilmark{1},
A. BUNKER\altaffilmark{5},
C. R. BRIDGE\altaffilmark{4},
and R. G. ABRAHAM\altaffilmark{4}
}

\altaffiltext{1}{Carnegie Observatories, 813 Santa Barbara St, Pasadena, 
CA 91101, U.S.A.}
\altaffiltext{2}{Center for Space Research, Massachusetts Institute of 
Technology, Cambridge, MA 02139-4307, U.S.A.}
\altaffiltext{3}{Department of Physics and Astronomy, San Francisco State 
University, San Francisco, CA 94132-4163, U.S.A.}
\altaffiltext{4}{Department of Astronomy, University of Toronto, Toronto ON, 
M5S~3H8 Canada}
\altaffiltext{5}{Institute of Astronomy, Cambridge CB3 OHA, England, UK}
\newpage

\begin{abstract}

  We present rest-frame $R$-band galaxy luminosity function measurements for 
three different redshift ranges: $0.5\leq z \leq 0.75$, $0.75 \leq z \leq 1.0$,
and $1.0\leq z \leq 1.5$.  Our measurements are based on photometric redshifts 
for $\sim 3000$ $H$-band selected galaxies with apparent magnitudes $17\leq H 
\leq 20$ from the Las Campanas Infrared Survey.  We show that our photometric 
redshifts are accurate with an RMS dispersion between the photometric and 
spectroscopic redshifts of $\sigma_z/(1+z) \approx 0.08$.  Using galaxies 
identified in the Hubble Deep Field South and Chandra Deep Field South regions,
we find, respectively, that ($7.3\pm 0.2$)\,\% and ($16.7\pm 0.4$)\,\% of the 
$H\leq 20$ galaxies are at $z\geq 1$.  We first demonstrate that the systematic
uncertainty inherent in the luminosity function measurements due to 
uncertainties in photometric redshifts is non-negligible and therefore must
be accounted for.  We then develop a technique to correct for this systematic 
error by incorporating the redshift error functions of individual galaxies in 
the luminosity function analysis.  The redshift error functions account for the
non-gaussian characteristics of photometric redshift uncertainties.  They are 
the products of a convolution between the corresponding redshift likelihood 
functions of individual galaxies and a Gaussian distribution function that 
characterizes template-mismatch variance.  We demonstrate, based on a Monte 
Carlo simulation, that we are able to completely recover the bright end of the 
intrinsic galaxy luminosity function using this technique.  Finally, we 
calculate the luminosity function separately for the total $H$-band selected 
sample and for a sub-sample of early-type galaxies that have a best-fit 
spectral type of E/S0 or Sab from the photometric redshift analysis.  The 
primary results of this analysis are: (1) the galaxy luminosity functions are 
consistent with a Schechter (1976) form; (2) the evolution of the co-moving 
luminosity density $\ell$ of $H$-band selected galaxies is characterized by 
$\Delta\log \ell\,/\Delta\log (1+z)=0.6 \pm 0.1$ at rest-frame 6800\,\AA; and
(3) $\ell_R$ for color-selected early-type galaxies exhibits moderate 
evolution from $z\sim 1.5$ to $z \sim 0.3$.  Specifically, $\ell_R$ for these
red galaxies brighter than 1.0 (1.6) $L_*$ could decrease with increasing 
redshift by {\em at most} a factor of three (six) over this redshift range 
after removing possible stellar fading according to the most extreme stellar 
evolution scenario.

\end{abstract}


\newpage

\section{Introduction}

  The galaxy luminosity function represents the product of a series of physical
processes that govern galaxy evolution, including star formation and the 
recycling of metals through various heating and cooling processes.  An accurate
estimate of the galaxy luminosity function as a function of time therefore 
bears significantly on our understanding of how structures in the universe form
and evolve.  Independent measurements of the optical galaxy luminosity function
based on various deep redshift surveys between $z\sim 0$ and $z\sim 0.75$ have
yielded consistent results (Lilly \etal\ 1995; Ellis \etal\ 1996; Marzke \& da 
Costa 1997; Bromley \etal\ 1998; Marzke \etal\ 1998; Lin \etal\ 1999; Madgwick 
\etal\ 2001).  For example, the galaxy luminosity function is well represented 
by a Schechter function (Schechter 1976) with a steeper faint-end slope for 
galaxies of bluer optical colors at all epochs that have been studied.  
Comparisons of these measurements at different redshifts further show that 
these blue galaxies exhibit a stronger evolution than red galaxies both in 
their space density and in their rest-frame optical luminosity.  

  It is difficult, however, to apply these results to fully constrain galaxy 
formation models.  While galaxy luminosities measured in the optical are 
sensitive to the transient star formation rate and therefore are a good tracer
of the instantaneous star formation activity, they are not representative of 
the total stellar mass that has been assembled in the galaxies over time.  
Additional caveats exist in all studies that attempt to draw conclusions based 
on comparisons of luminosity function measurements obtained using different 
galaxy samples.  For example, comparison of morphologically-selected and 
color-selected galaxy samples is inadequate, because even morphologically 
distinct galaxies can exhibit similar colors.  Furthermore, it is impossible to
unambiguously identify the same galaxy population at different epochs using the
observed colors because the underlying stellar evolution sequence is unknown.  
Finally, other selection biases such as the cosmic surface brightness dimming 
effect (preventing accurate morphological classifications at high redshift), 
bandpass selection effect (galaxies selected in the same observed-frame 
bandpass may have different rest-frame properties), and the presence of dust 
(in which a young, star-forming galaxy masquerades as an old, quiescent one) 
further complicate the interpretations of comparisons of different galaxy 
surveys.

  We have initiated the Las Campanas Infrared (LCIR) survey (Marzke \etal\ 
1999; McCarthy \etal\ 2001; Chen \etal\ 2002; Firth \etal\ 2002) to study the 
history of mass assembly based on a uniform sample of near-infrared selected 
galaxies.  The rest-frame near-infrared light is a good proxy for the total 
stellar mass, independent of age or galaxy type (Charlot 1996).  Measurements 
of the galaxy luminosity function at rest-frame near-infrared wavelengths at 
different epochs may, therefore, be adopted to estimate the stellar mass 
density evolution.  A robust measurement of the rest-frame near-infrared 
luminosity function at $z \approx 1$ requires a galaxy survey conducted in the 
$K$ band.  We have so far completed the LCIR $H$-band survey and a complete 
$K$-band survey is underway.  Here we present measurements of the rest-frame 
$R$-band luminosity functions in three redshift intervals over the range 
$z=0.5$--1.5 for the $H$-band selected galaxies using photometric redshifts.
At $z\approx 1.2$, the observed-frame $H$-band approximately corresponds to 
rest-frame $R$-band, and therefore the calculation does not depend sensitively
on the templates adopted to determine the $k$-correction.  The primary 
objectives of the analysis are (1) to obtain measurements of the galaxy 
luminosity function at $z\geq 1$, where different galaxy formation scenarios 
begin to show distinct predictions in the space densities and masses of evolved
galaxies, and (2) to study galaxy evolution at redshifts between $z\approx 0.5$
and $z\approx 1.5$ based on a uniform sample of near-infrared selected 
galaxies, thus extending the redshift range of existing deep redshift surveys 
to beyond $z=1$.

  We first conducted a photometric redshift survey in the LCIRS galaxy sample 
to identify a complete sample of galaxies at $z>0.5$.  The photometric redshift
survey in two of the survey fields identified $\geq 3000$ galaxies at $z\geq 
0.5$, yielding a statistically significant sample of intermediate-redshift 
galaxies for understanding the galaxy population at redshifts beyond $z=1$.  
The sample size is comparable to those of wide-field spectroscopic surveys 
carried out at $z\leq 0.5$ such as the Canadian Network for Observational 
Cosmology Field Galaxy Redshift Survey (CNOC2; e.g.\ Lin \etal\ 1999).  The 
photometric redshift techniques determine galaxy redshifts based on the 
presence/absence of broad-band spectral discontinuities, rather than 
narrow-band emission or absorption line features.  They are therefore of 
particular value for the identification of galaxies at $z\geq 0.75$.  Complete 
{\em spectroscopic} surveys of faint galaxies in this redshift range are 
extremely challenging, as prominent spectral line features in galaxy spectra 
are redshifted to near-infrared wavelengths and are difficult to detect due to 
bright, contaminating sky lines and detector limitations.  It has been 
demonstrated over the past several years that distant galaxies may be robustly
identified using photometric redshift techniques that incorporate optical and 
near-infrared broad-band photometry (Connolly \etal\ 1997; Lanzetta, 
Fern\'{a}ndez-Soto, \& Yahil 1998; Fern\'{a}ndez-Soto, Lanzetta, \& Yahil 1999;
Martini 2001; Fern\'{a}ndez-Soto \etal\ 2001; Rudnick \etal\ 2001).  In this 
paper, we show that the photometric redshift measurements of the galaxies 
identified in the LCIR survey are accurate, with an RMS dispersion between the
photometric and spectroscopic redshifts of $\sigma_z/(1+z) \approx 0.08$. 

  Next, we developed a maximum likelihood method to measure galaxy luminosity
function using photometric redshifts, which explicitly accounts for photometric
redshift uncertainties of individual galaxies.  Photometric redshift 
uncertainties together with a steep slope of the galaxy luminosity function 
would flatten the observed luminosity function and yield systematic biases.  
Uncertainties in photometric redshifts must therefore be accounted for in the
luminosity function analysis.  We demonstrate that systematic uncertainties in
the derived galaxy luminosity function are significantly reduced when redshift
error functions are taken into account.  We show that (1) the galaxy luminosity
function is consistent with a Schechter (1976) form; (2) the evolution of the 
co-moving luminosity density of the $H$-band selected galaxy is characterized 
by $\Delta \log \ell\,/\Delta\log (1+z)=0.6 \pm 0.1$ at rest-frame 6800\,\AA; 
(3) galaxies with spectral energy distributions best characterized as E/S0 or 
Sab appear to have faded by $\approx 0.5$ mag from $z \sim 1$ to $z\sim 0.5$, 
consistent with the expected luminosity evolution from an exponentially 
declining star formation rate model; and (4) only moderate evolution in the 
space density and in the rest-frame, co-moving $R$-band luminosity density is 
found for the bright, color-selected early-type galaxies.

  We adopt the following cosmology: $\Omega_{\rm M}=0.3$ and $\Omega_\Lambda=
0.7$ with a dimensionless Hubble constant $h = H_0/(100 \ {\rm km} \ {\rm 
s}^{-1}\ {\rm Mpc}^{-1})$ throughout the paper.  

\section{The Las Campanas Infrared Survey}

  The LCIR survey is a deep, wide-field near-infrared and optical imaging 
survey designed to identify a large number of galaxies dominated by an old 
stellar population at $z\geq 1$, while securing a uniform sample of galaxies of
all types to $z\sim 2$ based on their broad-band optical and near-infrared 
colors.  We have completed the $H$-band survey, covering one square degree of 
sky to a mean $5\,\sigma$ detection limit in a four arcsecond diameter aperture
of $H \sim 20.8$.  Initial catalogs of $\apg$ 54,000 galaxies identified over 
1400 ${\rm arcmin}^2$ together with results from a comprehensive study of the 
survey incompleteness in each field are presented in Chen \etal\ (2002).  The 
analysis presented in this paper is based on galaxies identified in two of the
LCIR fields: the Hubble Deep Field South (HDFS) and Chandra Deep Field South 
(CDFS).  To summarize, the galaxy sample contains $\sim$ 6,700 $H$-band 
selected galaxies over 847 ${\rm arcmin}^2$ in the HDFS region with 
complementary optical $U$, $B$, $V$, $R$, and $I$ colors, and $\sim 7,400$ 
$H$-band selected galaxies over 561 ${\rm arcmin}^2$ in the CDFS region with 
complementary optical $V$, $R$, $I$, and $z'$ colors.  

  To study the statistical properties of galaxies at $z\geq 1$ using the LCIR
survey sample, we need to first determine redshifts for the galaxies.  The
photometric redshift techniques that determine galaxy redshifts based on their
broad-band spectral energy distribution (SED) offer an efficient means to 
obtain robust redshift measurements together with well-understood measurement 
uncertainties for all galaxies in the survey.  

\subsection{Photometric Redshift Likelihood Analysis}

  We adopted the photometric redshift technique originally developed by 
Lanzetta and collaborators (e.g.\ Lanzetta, Yahil, \& Fern\'{a}ndez-Soto 1996),
and modified the program for its application to ground-based imaging data.  The
technique determines galaxy redshifts by comparing the SED of a galaxy, which 
is established based on a compilation of broad-band photometric measurements, 
with a grid of spectrophotometric templates evaluated at different redshifts. 
It employs a maximum likelihood method, in which the likelihood of a galaxy 
matching a given spectral type $T$ at redshift $z$ based on $n$ photometric 
measurements is written as
\begin{equation}
{\cal L}_z(z,T) = \prod_{i=1}^n \exp\left\{-\frac{1}{2}\left[\frac{f_i-AF_i(z,T)}{\sigma_i}\right]^2\right\},
\end{equation}
where $f_i$ and $F_i$ are respectively the measured and model galaxy fluxes in
bandpass $i$, $A$ is the normalization of the template, and $\sigma_i$ is the
measurement uncertainty of $f_i$.  In cases where the galaxy is not observed in
$m$ of the $n$ bandpasses to the detection limits of the images, equation (1) 
is modified as
\begin{equation}
{\cal L}_z(z,T) = \left ( \prod_{i=1}^{n-m} \exp\left\{-\frac{1}{2}\left[\frac{f_i-AF_i(z,T)}{\sigma_i}\right]^2\right\} \right ) \left ( \prod_{i=1}^{m} \int_{f_{\rm min}}^{f_{\rm max}} d\,f'\exp\left\{-\frac{1}{2}\left[\frac{f'-AF_i(z,T)}{\sigma_i}\right]^2\right\} \right ),
\end{equation}
where $f_{\rm min}$ and $f_{\rm max}$ are the minimum and maximum possible 
observed fluxes of the galaxy.  We have chosen $f_{\rm min}=0$ and $f_{\rm max}
=2\sigma_i$.  The likelihood function ${\cal L}_z(z,T)$ is maximized with 
respect to the type $T$ at different redshifts to form ${\cal L}_z(z)$, which 
is then maximized with respect to the redshift $z$ to determine a best-fit 
photometric redshift $z_{\rm phot}$ for the galaxy.  

\subsection{Spectral Energy Distribution Templates}

  We included in our photometric redshift likelihood analysis six galaxy 
templates, a suite of stellar templates, and a QSO template in order to 
effectively identify the large number of stars and potential QSO/AGN candidates
that appear in wide-field surveys.  Every template covers a spectral range from
$\lambda = 300$ to 25,000 \AA.  The galaxy templates were adopted from 
Fern\'andez-Soto \etal\ (1999) and Yahata \etal\ (2000) and spanned a range of
spectral types---from elliptical or S0 (E/S0), Sab, Scd, irregular (Irr), to 
starburst (SB).  A total of 150 stellar templates were compiled from the 
literature, spanning a wide range of stellar types---from early-type O and B 
stars to late-type L and T dwarfs (Oppenheimer \etal\ 1998; Pickles 1998; Fan 
\etal\ 2000; Leggett \etal\ 2000).  The QSO template was formed by first 
adopting the composite QSO spectrum presented in Francis \etal\ (1991) and 
extrapolating the spectrum to infrared wavelengths using a simple power law. 
Because the Francis spectrum was derived based in part on a large number of 
QSOs at $z\geq 2$ and was not corrected for intervening absorption at 
wavelengths shortward of the redshifted \lya\ emission line, we replace the 
continua at $\lambda \leq 1450$ \AA\ with the composite spectrum presented in 
Zheng \etal\ (1997) based on $\sim 100$ spectra of QSOs at $z\leq 1.6$ obtained
with the Faint Object Spectrograph aboard the Hubble Space Telescope.  Our 
approach to separate stars from galaxies is described in Chen \etal\ (2002). 
The number-magnitude diagram of the stars identified in the HDFS and CDFS 
fields, in comparison to that of the galaxies, is presented in Chen \etal\ as 
well.

\subsection{The Photometric Redshift Survey}

  We performed the redshift likelihood analysis and determined photometric 
redshifts for all galaxies identified in the LCIRsurvey using the available 
near-infrared and optical broad-band photometry.  As discussed in \S\ 2.4, 
uncertainties in galaxy photometry contribute to the uncertainties in 
photometric redshifts.  Because galaxies vary in size, fixed-aperture 
photometry would include excess noise from sky for smaller objects and exclude 
flux from the outskirts of bigger objects, both of which would yield a reduced 
signal-to-noise (S/N) ratio in the photometric measurements.  Fixed-aperture 
photometry is, therefore, not optimal for establishing the observed SEDs of 
individual galaxies to be compared with model templates in the photometric
redshift analysis.  

  It is clear that to improve the accuracy of photometric redshifts, we need to
first achieve an optimal S/N ratio for the photometric measurements of 
individual galaxies in all bandpasses, where by optimal measurements we mean
those that do not include excess noise or exclude object fluxes.  To reach the
goal, we obtained a new set of photometric measurements using a varying-size 
aperture for different objects, which is different from the fix-aperture 
photometry published in Chen \etal\ (2000).  The individual object apertures 
were adopted from the object segmentation determined by applying SExtractor 
(Bertin \& Arnout 1996) to the ``white light'' image of each field.  As 
described in Chen \etal\ (2002), the white light image is an unweighted sum of
all the registered optical images that had been scaled to unit exposure time 
and filter throughput.  Because of the improved signal, particularly in the low
surface brightness regions, this image is more sensitive than any individual 
optical images for SExtractor to identify the true extent of each object at a 
given surface brightness detection threshold.  We measured, for each galaxy, 
the total fluxes and associated uncertainties in different bandpasses using the
same aperture determined in the white light image by SExtractor.  This 
procedure allowed us to recover the total fluxes of each object in different 
bandpasses, without significantly compromising the precision of the photometric
measurements.

  Note that the main objective of the analysis here is to obtain accurate and
precise broad-band colors.  Accurate measurements of the total fluxes of 
individual galaxies are necessary for a robust estimate of the galaxy 
luminosity function, but not for the purpose of obtaining an accurate 
photometric redshift.  We adopted the optimal photometric measurements for
the photometric redshift analysis and the 4''-aperture photometric measurements
for the luminosity function analysis.  We present in Figure 1 the median 
residuals and the corresponding $1\,\sigma$ scatters between the ``optimal'' 
and 4''-aperture photometric measurements versus magnitude for the $H$-band 
selected objects in the CDFS region in the $V$, $R$, $I$, and $H$ bands.  It 
was shown in Chen \etal\ (2002) that the 4''-aperture photometric measurements
closely resembled the total magnitudes of individual galaxies---only $< 0.1$ 
mag aperture correction was necessary.  Figure 1 shows that the optimal 
photometry is consistent with the fixed aperture photometry in all bandpasses 
to within measurement uncertainties, although it systematically underestimates
the total fluxes at faint limits.  We note that the systematic biases of the 
optimal photometry at faint magnitudes do not affect our photometric redshift 
estimates, because they are almost entirely independent of the adopted
bandpasses.

  The products of the photometric redshift analysis for each galaxy include: 
the photometric redshift measurement, the redshift likelihood function, and the
best-fit spectral template.  Figure 2 shows examples of the results in pairs of
panels for four HDFS galaxies that have been spectroscopically identified at
different redshifts.  The top panel of each pair shows the photometric 
measurements and uncertainties as solid points with error bars.  The solid 
curves represent the best-fit spectral templates with the spectral type 
indicated in the upper-left corner.  The open squares indicate the predicted 
fluxes in individual bandpasses, estimated by integrating the best-fit template
over the corresponding system transmission functions. The dashed lines indicate
the zero flux level. The bottom panel of each pair shows the redshift 
likelihood functions.  The estimated photometric redshift ($z_{\rm phot}$) and 
spectroscopic redshift ($z_{\rm spec}$) are indicated in the lower-right corner
of each of the top panels.  In all four cases, we have successfully determined 
the galaxy redshifts to within the estimated uncertainties.

  We studied the redshift distribution of the $H$-band selected sample. The 
left panel of Figure 3 shows the redshift distribution of the $H$-band selected
HDFS galaxies with $H \leq 20$.  The photometric redshifts were obtained based 
on the available $U$, $B$, $V$, $R$, $I$, and $H$ photometric measurements. 
The hatched region indicates the redshift distribution of galaxies with $I-H
\geq 3$.  The $I-H\geq 3$ color criterion is predicted by various evolutionary
scenarios to identify evolved galaxies at $z \geq 1$ (Chen \etal\ 2002).  The 
curves indicate the predicted redshift distributions of the total $H$-selected
sample and the red population, which were calculated using the model that best 
fit the number--magnitude relation presented in Chen \etal\ (2002).  
Specifically, we calculated the predicted redshift distribution of the total 
$H$-band selected galaxies by adopting the $K$-band luminosity function 
obtained by Gardner \etal\ (1997), assuming a rest-frame color of $H-K\sim 0.2$
(e.g.\ Mobasher \etal\ 1986) and a no-evolution scenario.  The predicted 
redshift distribution of the red galaxies was estimated based on the observed
population fraction of elliptical galaxies in the local universe (e.g.\ 
Binggeli, Sandage, \& Tammann 1988).  We first adopted the same luminosity 
function and scaled it according to $M_*({\rm red})=M_*-1.2$ and $\phi_*({\rm 
red})=0.2\,\phi_*$ for identifying the red population.  Next, we assumed an 
exponentially declining star formation rate history with a 1-Gyr e-folding time
for galaxies formed at $z_f=30$ to model the pure luminosity evolution and 
added random photometric noise of $\sim 0.25$ mag in the model to account for 
the uncertainty of the $I-H$ color selection criterion in the data.  Finally,
we calculated the integrated galaxy number counts versus redshift.  Figure 3 
shows that these models agree well with the observation.

  In the right panel of Figure 3, we present the cumulative redshift 
distributions of the $H\leq 20$ galaxies in the HDFS (open) and CDFS (hatched
histogram) regions.  We exclude the CDFS galaxies identified at $z\leq 0.75$ 
from our analysis as they lack $U$- and $B$-band data, which causes photometric
redshifts to be less accurate in this range.  We find that ($7.3\pm 0.2$)\,\% 
(HDFS) and ($16.7\pm 0.4$)\,\% (CDFS) of the $H$-band selected galaxies with 
$H \leq 20$ are at $z\geq 1$.  The errors were estimated using a Monte Carlo 
simulation, which takes into account photometric uncertainties and photometric
redshift uncertainties, the subject of the next section.  The difference 
between the two fields in the number density of $z>1$ galaxies is likely due to
the fluctuations in large-scale structures.  In particular, our survey is
sensitive to galaxies at the bright end of the galaxy luminosity function, 
which are believed to be more strongly clustered.

\subsection{Uncertainties in Photometric Redshift Measurements}

  In order for any statistical analysis of the intrinsic properties of galaxies
identified using photometric redshift techniques to be plausible, we must first
quantify the uncertainties.  As discussed in detail in Lanzetta \etal\ (1998) 
and Fern\'andez-Soto \etal\ (2001, 2002), uncertainties in photometric redshift
measurements are due to two sources: (1) for bright galaxies, the dominant 
uncertainty is due to the finite number of templates employed in the analysis; 
this is known as template-mismatch variance, and (2) for faint galaxies, an 
additional uncertainty is due to photometric measurement uncertainties.  
Fern\'andez-Soto \etal\ (2001) demonstrated that uncertainties in photometric
redshifts due to template mismatch is well characterized by a Gaussian 
distribution function with the 1-$\sigma$ width equal to the RMS residuals 
between photometric and spectroscopic redshifts.  In addition, equation (1) 
explicitly shows that in the absence of template-mismatch variance the 
uncertainty of the redshift likelihood function depends entirely upon the 
accuracy and precision of the flux measurements.  Namely, the smaller the 
$\sigma_i$'s are, the more precise the best-fit photometric redshift 
measurement.  Uncertainties in photometric redshifts due to photometric errors
are, therefore, well represented by the redshift likelihood functions of 
individual objects.  

  Photometric redshift uncertainties can be quantified by explicitly taking 
into account the redshift likelihood functions without adopting a parametric 
form.  The error function of the photometric redshift measurement of galaxy $i$
is 
\begin{equation}
p_i(z-z_i;z_i) = \int_0^\infty\frac{1}{\sigma_z\sqrt{2\,\pi}}\,{\cal L}_z^i(z'-z_i;z_i)\cdot\exp\,[-(z-z')^2/2\,\sigma_z^2]\,d\,z'.
\end{equation}
Equation (3) is a convolution of the redshift likelihood function 
${\cal L}_z^i(z)$ and a Gaussian kernel of width $\sigma_z=\hat\sigma_z(1+z_i)$
that characterizes the template-mismatch variance, where $\hat\sigma_z$ is the
RMS residuals between spectroscopic and photometric redshifts at zero redshift.

  To assess the accuracy and reliability of our photometric redshifts, we first
compared our photometric redshifts with known spectroscopic redshifts.  There
are over 100 objects in the HDFS region (Glazebrook 1998; Tresse \etal\ 1999; 
Palunas \etal\ 2000; Cristiani \etal\ 2000; Dennefeld 2002).  Excluding the 
ambiguous spectroscopic redshift identifications indicated by these authors, we
found 67 objects that have both reliable spectroscopic and photometric redshift
measurements at $z \leq 2.5$, 13 of which are stars.  There are 162 known
spectroscopic redshifts in the CDFS region obtained by ourselves (McCarthy 
\etal\ 2002, in preparation).  Excluding ambiguous photometric redshifts 
(objects with uncertain redshift likelihood distributions), we found 140 
objects with both reliable spectroscopic and photometric redshifts.

  Figure 4 shows the comparisons of photometric ($z_{\rm phot}$) and 
spectroscopic ($z_{\rm spec}$) redshifts for the HDFS and CDFS objects in the 
left panel.  Solid points represent objects identified in the HDFS region and 
open circles represent those in the CDFS region.  The solid line indicates 
$z_{\rm phot}=z_{\rm spec}$.  The distributions of redshift residuals are shown
in the right panel of Figure 4 for the HDFS (open histogram) and CDFS (shaded 
histogram) objects.  We find an RMS dispersion of $\sigma_z/(1+z) \approx 0.08$
for the 67 HDFS objects at $z \leq 1$.  Only one of the 13 stars was 
misidentified as a galaxy at $z_{\rm phot} = 0.09$.  Note that the mean
redshift residuals are nearly zero at all redshifts that have been tested here,
supporting that there is no systematic bias in our photometric redshifts.  It 
demonstrates that the photometric redshifts obtained using $UBVRIH$ photometry 
are accurate at all redshifts below $z\approx 1$.  The comparison based on the 
34 CDFS objects at $z \geq 0.75$ shows $\sigma_z/(1+z) \approx 0.08$.  It 
demonstrates that the photometric redshifts obtained using $VRIz'H$ photometry 
are accurate at redshifts beyond $z\sim 0.75$, although they become unreliable 
at lower redshifts due to the lack of $U$ and $B$ photometry.

  There are only a small number of known spectroscopic redshifts at $1\leq z 
\leq 2.5$ in both the HDFS and CDFS regions, but it is unlikely that the 
accuracy of the photometric redshifts at $1\leq z\leq 2.5$ would be much worse
than those at lower redshifts.  It has been demonstrated by different groups 
that photometric redshifts are accurate to within $\sigma_z/(1+z) < 0.1$ at 
$1\leq z\leq 1.5$ for all galaxies that have been tested so far (Connolly 
\etal\ 1997; Fern\'andez-Soto \etal\ 2001; Rudnick \etal\ 2001).  In 
particular, Fern\'andez-Soto \etal\ (2001) showed that using the same set of 
SED templates their photometric redshifts are consistent with spectroscopic 
redshifts for all 19 galaxies that have known spectroscopic redshifts at $1
\leq z\leq 1.5$ in the HDF north.  This supports that the number of templates 
in our photometric redshift analysis is sufficient to minimize template 
mismatch uncertainties.  A more critical examination of the accuracy of our 
photometric redshifts in this redshift range relies on the availability of 
more spectroscopic measurements.

  Next, we performed a series of simulations designed to evaluate the 
uncertainties in the photometric redshifts due to purely photometric 
uncertainties.  We first generated a set of photometric measurements in each 
bandpass for a given $H$-band magnitude and input redshift $z_{\rm in}$ based 
on an E/S0 template.  We then perturbed the photometric measurements in 
individual bandpasses within the $1\,\sigma$ photometric uncertainties measured
for the individual optical and $H$ images.  Next, we determined a best-fit 
redshift $z_{\rm out}$ using the photometric redshift technique.  Finally, we 
repeated this procedure 1000 times and calculated the histogram of the 
residuals between $z_{\rm in}$ and  $z_{\rm out}$.  We repeated the entire 
process for the Sab, Scd, and Irr spectral templates to assess the performance
of the photometric redshift technique for galaxies of different spectral shape.
Figure 5 shows an example of the simulation results based on the filter 
combination and image sensitivities of the HDFS data.  The left panels show the
histograms of the residuals between the input galaxy redshift $z_{\rm in}$ and
the best-fit photometric redshift $z_{\rm out}$ and the right panels show the 
cumulative fractions versus absolute redshift residuals, $\Delta z = |z_{\rm 
out}-z_{\rm in}|$.  The results indicate that while we could accurately 
identify an Sab-Irr galaxy of $H=20$ at $z=1.2$, there is an $\sim 10\,$\% 
chance that we would misidentify an E/S0 galaxy of $H=20$ at $z=1.2$ as a 
starburst galaxy at $z\sim 5$.  The bimodal distribution of the residuals is 
due to the confusion between a 4000-\AA\ spectral discontinuity at low 
redshifts and a Ly$\alpha$ continuum break at high redshifts.

  We repeated the simulations for a grid of redshift and $H$-band magnitude
based on the filter combination and image sensitivities of both HDFS and CDFS
data.  The results are presented in Figure 6, each panel of which now shows 
the sum of the residual histograms over all four spectral types.  Figure 6 
indicates that, even in the absence of template-mismatch variance, photometric 
redshifts become progressively more uncertain at fainter apparent magnitudes 
and higher redshifts, as indicated by the broader distribution of the 
residuals.  The bimodal distribution that appears, for example, for an E/S0 
galaxy of $H=20$ at $z=1.2$ in the HDFS region does not exist for a galaxy of 
the same brightness at the same redshift range in the CDFS region, because the
CDFS images reach to fainter magnitude limits.  It is also clear, however, that
there exists a non-negligible probability that galaxies at $z\leq 0.4$ in the 
CDFS region would be misidentified as galaxies at $z\sim 3$ even at bright
magnitudes.  The misidentifications occur primarily in cases when the input 
spectral template was an early-type galaxy.  Because of the lack of $U$- and 
$B$-band photometric measurements for galaxies in the CDFS region, the observed
SEDs are poorly constrained at rest-frame wavelengths shortward of the 
4000-\AA\ break.  The ambiguity disappears for the CDFS galaxies at $z\geq 
0.8$, where the available $V$-band photometry begins to impose a strong 
constraint on the shape of the observed SEDs.  

  In Figure 7, we present the cumulative fractions of the simulated photometric
redshifts versus absolute redshift residuals for the four $z=1.2$ cases
presented in Figure 6.  The left panels show the simulation results for the 
HDFS galaxies and the right panels for the CDFS galaxies.  At $z=1.2$, it is 
clear that photometric redshifts are accurate to within $\Delta z = 0.1$ for 
the CDFS galaxies of $H\leq 21$, and that they become very uncertain for the 
HDFS galaxies at $H\sim 21$ because of their relatively larger photometric 
uncertainties.  Specifically, there is a $\sim$ 30\% chance that the 
photometric redshifts of these galaxies are off by more than $\Delta z = 0.1$.

  In summary, we conclude that the photometric redshift measurements of the 
HDFS objects based on $U$, $B$, $V$, $R$, $I$, and $H$ photometry are robust 
for galaxies with $H\leq 19$ at all redshifts $z\leq 1$.  The photometric 
redshift measurements of the CDFS objects based on $V$, $R$, $I$, $z'$, and $H$
photometry are robust for galaxies of $H\leq 20$ at all redshifts $0.75\leq z
\apl 1$.  We exclude CDFS objects with photometric redshifts $z\leq 0.75$ from
subsequent analysis because of the lack of photometric measurements in the 
observed $U$ and $B$ bands (see also Connolly \etal\ 1995).  The robustness of
photometric redshifts at $z\geq 1$ remains to be examined more critically, 
because only a small number of known spectroscopic redshifts are available in 
our survey fields.  However, consistent redshift estimates of the HDF galaxies
at $1\leq z \leq 1.5$ support the photometric redshift technique using the six
templates in our analysis for this redshift interval (e.g.\ Fern\'andez-Soto 
\etal\ 2001).  We note that redshift uncertainties due to dust reddening is 
not explicitly accounted for in our analysis.  We consider the effect of dust 
as part of sources that lead to template mismatch.  To further improve the 
reliability of the photometric redshift measurements and to reduce the number 
of low-redshift galaxies that are mis-identified as star-forming galaxies at 
high redshifts due to photometric uncertainties, we restricted the redshift 
range of the photometric redshift likelihood analysis to $z\leq 2.5$.  This is
appropriate because the fraction of $H\leq 20$ galaxies at $z\geq 2.5$ is 
$\sim\,0.1$\,\% (Lanzetta \etal\ 1999; Cohen \etal\ 2000; Shapley \etal\ 2001).

\section{Rest-frame $R$-band Galaxy Luminosity Function}

   Here we present the rest-frame $R$-band luminosity function for three 
redshift ranges: $0.5\leq z \leq 0.75$, $0.75 \leq z \leq 1.0$, and $1.0\leq z
\leq 1.5$.  We emphasize that the results are based on {\em photometric 
redshifts} for $\sim 3000$ $H$-band selected galaxies with apparent magnitude 
$17 \leq H \leq 20$.  We calculated the luminosity function separately for the 
total $H$-band selected sample and for a sub-sample of early-type galaxies that
have a best-fit spectral type of E/S0 or Sab in the photometric redshift 
analysis.  The primary objectives are (1) to obtain an estimate of the galaxy 
luminosity function at $z\sim 1$ in order to study the statistical properties 
of galaxies at this epoch, and (2) to study galaxy evolution at redshifts 
$ 0.5 \leq z \leq 1.5$ based on a uniform sample of near-infrared selected 
galaxies; this extends the redshift range of existing deep redshift surveys to
beyond $z=0.75$. 

  The absolute $R$-band magnitude $M_R$ of a galaxy with an apparent magnitude 
$H$ at redshift $z$ is 
\begin{equation}
M_R(z) = H -25.0 - 5\times\log\,\frac{D_L(z)}{(\mbox{Mpc})}-k'(z) + 2.5\times\log (1+z),
\end{equation}
where $D_L(z)$ is the luminosity distance to the galaxy in units of Mpc and 
$k'(z)$ is the $k$-correction term to account for the color difference between
the observed-frame and corresponding rest-frame bandpasses.  The $k$-correction
term for each galaxy was calculated using the best-fit spectral template from 
the photometric redshift analysis.  For galaxies at $z\sim 1.2$, the 
observed-frame $H$ band approximately corresponds to rest-frame $R$, so the 
calculation does not depend sensitively upon the adopted templates to determine
the $k$-correction.  For galaxies at $z < 1$, the estimated $k$-correction 
ranges from between $-0.34$ (SB) and $-0.9$ (E/S0) at $z=0.5$ to between 
$-0.12$ (SB) and $-0.43$ (E/S0) at $z=1.0$.

  Several methods to calculate the luminosity function have been proposed over
the years.  We examine three of these: (1) the $1/V_{\rm max}$ approach 
(Schmidt 1968; Felten 1976), (2) the Sandage-Tammann-Yahil (STY) approach 
(Sandage, Tammann, \& Yahil 1979), and (3) the stepwise maximum likelihood 
(SWML) approach (Efstathiou, Ellis, \& Peterson 1988) in the context of a 
photometric-redshift determination of the luminosity function.  A foreseeable 
systematic error in the derived luminosity function using photometric redshifts
is to overestimate the number density of galaxies at the bright end, producing
a flattened, deduced galaxy luminosity function relative to the intrinsic 
luminosity function.  This is due to the relatively large uncertainties 
associated with photometric redshifts and the steep slope at the bright end of
the galaxy luminosity function.  There are more intrinsically fainter galaxies 
scattered into the brighter end of the luminosity function than intrinsically 
brighter galaxies scattered into the fainter end, thus yielding an apparently 
flatter luminosity function.  To obtain robust estimates of the ``intrinsic'' 
galaxy luminosity function using photometric redshifts, it is therefore 
important to account for these systematic biases. In the following sections we 
present a maximum likelihood analysis that explicitly takes into account the 
empirical redshift error functions of individual objects in the luminosity 
function calculation, and demonstrate, via a Monte Carlo simulation, that our 
approach allows us to reduce the systematic errors and to completely recover 
the bright end of the galaxy luminosity function.  We summarize these three 
different luminosity function estimators, describe how we quantify and 
incorporate the photometric redshift uncertainties.

\subsection{The $1/V_{\rm max}$ Approach}

  The $1/V_{\rm max}$ approach was first proposed by Schmidt (1968) and 
modified by Felten (1976). The galaxy luminosity function $\phi(M)$
is related to the number of galaxies per co-moving volume expected in a survey 
defined by the limiting apparent magnitude range $m_{\rm min} \leq m \leq 
m_{\rm max}$ and the redshift range $z_{\rm min} \leq z \leq z_{\rm max}$ 
according to
\begin{equation}
\Delta n = \phi(M) (M_{\rm max} - M_{\rm min}) 
	 = \sum_i\frac{1}{V_i},\hspace{0.25in} \mbox{for $M_{\rm min} \leq M_i \leq M_{\rm max}$ and $z_{\rm min} \leq z_i \leq z_{\rm max}$}.
\end{equation}
The absolute magnitude range $[M_{\rm min},M_{\rm max}]$ corresponds to the 
apparent magnitude range imposed by the survey at the galaxy redshift $z_i$. 
The co-moving volume $V_i$ is the maximum accessible volume of the survey for a
galaxy of $M_i$,
\begin{equation}
V_i=\int_{z'_{\rm min}}^{z'_{\rm max}} \frac{c \Theta_i}{H_0} \frac{D_L^2}{(1+z)^2}\frac{dz}{\sqrt{\Omega_M(1+z)^3-(\Omega_M+\Omega_\Lambda-1)(1+z)^2+\Omega_\Lambda}},
\end{equation}
where $c$ is the speed of light, $\Theta_i$ is the angular area covered by the 
survey that is sensitive enough to detect the galaxy $M_i$ at $z_i$, and 
$z'_{\rm min}$ and $z'_{\rm max}$ are the minimum and maximum redshifts that 
satisfy both the limiting magnitude and redshift ranges of the survey.  The 
angular area $A_i$ for each galaxy identified in the LCIR survey is 
determined from the results of the incompleteness analysis presented in Chen 
\etal\ (2002).

  The $1/V_{\rm max}$ approach is essentially a maximum likelihood method that
does not assume any parametric form for estimating the galaxy luminosity 
function (see also \S\ 3.3 below).  It has been shown, however, that 
large-scale fluctuations in the galaxy number density introduce systematic
biases in the galaxy luminosity function obtained with this approach for a 
magnitude-limited sample (de Lapparent, Geller, \& Huchra 1989).  We show in 
\S\ 3.4 that it is difficult to recover the intrinsic galaxy luminosity 
function with this approach for a galaxy sample identified in a photometric 
redshift survey.

\subsection{The Sandage-Tammann-Yahil Approach}

  Previous studies have demonstrated that the galaxy luminosity function may be
well represented by a Schechter (1976) function,  
\begin{equation}
\Phi(L;L_*) = \phi_*\cdot (L/L_*)^{\alpha}\cdot\exp(-L/L_*)  
\end{equation}
or
\begin{equation}
\Phi(M;M_*) = (0.4 \ln 10)\cdot\phi_*\cdot 10^{0.4\,(M_*-M)(1+\alpha)}\cdot\exp(-10^{0.4\,(M_*-M)}).
\end{equation}
To accurately determine the faint-end slope $\alpha$ and $M_*$ for the
Schechter luminosity function in the presence of large-scale density 
fluctuations, Sandage, Tammann, \& Yahil (1979) first introduced a maximum 
likelihood approach that calculates the cumulative probability of observing an
ensemble of galaxies in a magnitude-limited survey for given $\alpha$ and 
$M_*$.  The probability of observing a galaxy of $M_i(m_i, z_i)$ in a 
magnitude-limited redshift survey given a Schechter luminosity function is
\begin{equation}
P_i(M_i;M_*) = \frac{10^{0.4\,(M_*-M_i(m_i,z_i))(1+\alpha)}\cdot\exp(-10^{0.4\,(M_*-M_i(m_i,z_i))})}{\Theta_i\cdot \int_{m_{\rm min}}^{m_{\rm max}} 10^{0.4\,(M_*-M(m,z))(1+\alpha)}\cdot\exp(-10^{0.4\,(M_*-M(m,z))}) \,dm},
\end{equation}
where $\Theta_i$ is the fraction of the angular area that is sensitive enough 
to detect the galaxy $m_i$ at redshift $z_i$ and $[m_{\rm min},m_{\rm max}]$ 
defines the magnitude sensitivity range of the survey.  If the luminosity 
function remains the same for galaxies with different clustering properties at 
different redshifts, then the likelihood of obtaining an ensemble of $N$ 
galaxies in the survey is
\begin{equation}
{\cal L}_\phi = \prod_{i=1}^{i=N}\,P_i(M_i;M_*). 
\end{equation}
The faint-end slope $\alpha$ and characteristic magnitude $M_*$ in equation (8)
are determined by maximizing the likelihood function ${\cal L}_\phi$.  

  The normalization $\phi_*$ of the Schechter luminosity function is determined
separately, according to
\begin{equation}
\phi_* = \sum_{i=1}^{i=N}\,V^{-1}\,\Bigg /\,\int_{M_{\rm lim}(z_i)}^\infty\bar{\Phi}(M;M_*)\,dM,
\end{equation}
where $V$ is the co-moving volume spanned from $z_{\rm min}$ to $z_{\rm max}$
in the survey, $\bar{\Phi}(M;M_*)$ is the galaxy luminosity function defined 
in equation (8) with $\phi_*=1$, and $M_{\rm lim}(z_i)$ is the corresponding 
absolute magnitude limit of the survey at the galaxy redshift $z_i$.  The 
uncertainty in $\phi_*$ may be estimated by bootstrap resampling, that is, 
randomly re-sampling the same number of galaxies from the parent catalog a 
large number of times (allowing for duplications) and measuring the variation 
of $\phi_*$ between different re-sampled catalogs.

\subsection{The Stepwise Maximum Likelihood Approach}

  A different maximum likelihood approach was developed by Efstathiou, Ellis, 
\& Peterson (1988), one which does not require a functional form in the 
analysis and is similarly insensitive to the presence of large-scale density 
fluctuations in the survey.  In this approach, the galaxy luminosity function 
is parameterized according to
\begin{equation}
\Phi(M) = \phi_k(M_k), \hspace{0.25in}\mbox{if $M_k-\frac{1}{2}\Delta M \leq M < M_k+\frac{1}{2}\Delta M$, for $k=1,..,L$}.
\end{equation}
The probability of observing a galaxy of $M_i(m_i,z_i)$, for which $M_k-
\frac{1}{2}\Delta M \leq M_i(m_i,z_i) < M_k+\frac{1}{2}\Delta M$ is, 
\begin{equation}
P_i(m_i,z_i) = \phi_k(m_i,z_i)\,\Bigg /\,\sum_{k=k(m_{\rm min})}^{k=k(m_{\rm max})}\,\phi_k(m,z_i)\,\Delta m,
\end{equation}
where $k(m_{\rm min},z_i)$ and $k(m_{\rm max},z_i)$ indicate the minimum and 
maximum absolute magnitude intervals defined by the apparent magnitude 
sensitivity range of the survey at $z_i$.  The likelihood of observing $N$ 
galaxies in a magnitude-limited survey is then
\begin{equation}
{\cal L}_\phi = \prod_{i=1}^{i=N}\,P_i(m_i,z_i).
\end{equation}
The best-fit $\phi_k$'s are obtained by maximizing the likelihood function 
defined in equation (14).  

  In the simplest case where $k(m_{\rm min})=k_0$ and $k(m_{\rm max})=k_f$ for 
$i = 1$--$N$, it is straightforward to show that $\phi_k \propto m_k$, where 
$m_k$ is the number of galaxies in each magnitude interval.  Therefore, in the
absence of large-scale density fluctuations, the SWML approach and the 
$1/V_{\rm max}$ approach are essentially the same.

\subsection{Systematic Uncertainties}

  Photometric redshift techniques clearly provide an efficient means of 
identifying distant, faint galaxies in deep, wide-field surveys.  It is, 
however, important to understand how the redshift measurement errors affect the
derived luminosity functions of distant galaxies.  Photometric redshift 
uncertainties propagate through the uncertainties in luminosity distances and 
consequently yield large uncertainties in the absolute magnitudes of individual
galaxies.  Here we neglect the effect of purely photometric uncertainties, 
because according to equation (4) a redshift uncertainty of $\delta z = 0.1$ 
will result in a magnitude uncertainty of $\Delta M \sim 0.4$ at $z\sim 1$ due
to the uncertainty in the luminosity distance.  It is clear that photometric 
redshift errors dominate the uncertainties in the absolute magnitude 
calculations of individual galaxies.  

  We performed a Monte Carlo simulation in order to understand the systematic 
uncertainties in the galaxy luminosity function calculations due to errors in
photometric redshifts.  First, we generated a model catalog of 8000 galaxies of
different brightness at random redshifts between $z=0.1$ and $z=1$ according to
an input Schechter luminosity function.  Next, we assumed a redshift error
function parameterized as a Gaussian distribution function of $1\,\sigma$ width
$0.15\,(1+z_i)$ and formed an ``observed'' redshift catalog by perturbing the 
input galaxy redshift $z_i$ within the redshift error function.  Finally, we 
determined the luminosity function for the galaxies at $0.5 \leq z \leq 0.8$ 
using the $1/V_{\rm max}$ and STY methods.

  Figure 8 shows the derived galaxy luminosity function based on the simulated
catalogs using different approaches.  The input galaxy luminosity function is
shown as the dash-dotted curve with the selected $M_*$ and $\alpha$ indicated 
as a star in the inset.  The solid circles represent measurements using the
``observed'' galaxy redshifts based on the $1/V_{\rm max}$ approach.  The error
bars are the associated $1\,\sigma$ uncertainties estimated with a bootstrap
re-sampling technique that takes into account both the sampling and redshift
uncertainties.  The dotted curve represents the best-fit Schechter luminosity
function to the ``observed'' redshift catalog.  The best-fit $M_*$ and
$\alpha$ correspond to the upper solid point in the inset with the 99\% 
uncertainties indicated by the dotted contour.

  It is clear from Figure 8 that large redshift errors together with the steep
slope at the bright end of the galaxy luminosity function tend to flatten 
the observed luminosity function and result here in a best-fit $M_*$ that is
0.8 mag brighter than the input value.  This confirms the statements above that
more intrinsically fainter galaxies are scattered into the brighter end of the
luminosity function than are intrinsically brighter galaxies scattered into the
fainter end.  This is equivalent to convolving the galaxy luminosity function 
with an error function in the galaxy absolute magnitudes induced by the 
redshift uncertainties.  Evidently, redshift measurement uncertainties must be
accounted for in order to obtain an accurate estimate of the intrinsic galaxy 
luminosity function.

\subsection{A Modified Maximum Likelihood Analysis}

  In order to reduce the aforementioned bias we modified the likelihood 
analysis to incorporate the error functions of photometric redshift 
measurements into the probability calculations.  Because objects are scattered
in redshift space due to the relatively large redshift measurement errors, it 
is necessary to rewrite equations (9) and (13).  The probability of observing a
galaxy of $M_i(m_i,z_i)$ in a magnitude-limited photometric redshift survey is
now the cumulative probability for a galaxy of apparent magnitude $m_i$ at $z'$
to be identified at $z_i$ according to the redshift error function $p_i(z_i-z;
z_i)$ defined in equation (3).  Namely, it is a convolution of equations (9) 
and (13), respectively, with the error function of the photometric redshift 
measurement of the galaxy.  Equation (9) defined for the STY approach is 
therefore 
\begin{equation}
P_i(m_i,z_i;M_*) = \frac{\int_{0}^{z_f}\,10^{0.4\,(M_*-M_i(m_i,z'))(1+\alpha)}\cdot\exp(-10^{0.4\,(M_*-M_i(m_i,z'))})\cdot p_i(z_i-z';z_i)\,dz'}{\Theta\cdot \int_{m_{\rm min}}^{m_{\rm max}}\int_{0}^{z_f} 10^{0.4\,(M_*-M(m,z'))(1+\alpha)}\cdot\exp(-10^{0.4\,(M_*-M(m,z'))})\cdot\,p_i(z_i-z';z_i)\,dz'\,dm},
\end{equation}
where $z_f$ is the redshift limit of the grid search in the photometric 
redshift analysis, and equation (13) defined for the SWML approach is
\begin{equation}
P_i(m_i,z_i) = \frac{\int_0^{z_f}\,\phi_k(m_i,z') \cdot p_i(z_i-z';z_i)\,dz'}{\sum_{k=k(m_{\rm min})}^{k=k(m_{\rm max})}\int_0^{z_f}\phi_k(m,z') \cdot p_i(z_i-z';z_i)\,dz'\Delta m}.
\end{equation}
In cases where an accurate and precise spectroscopic redshift measurement is 
available, $p_i(z_i-z;z_i)=\delta(z-z_i)$ and equations (15) and (16) reduce 
to equations (9) and (13).  The parameters that determine the Schechter 
luminosity function, $\alpha$ and $M_*$ in the STY approach and the $\phi_k$'s
in the SWML approach, are then evaluated with the maximum likelihood functions
(equations 10 and 14).

  We applied this modified likelihood analysis to the simulated catalog 
described in \S\ 3.4 in order to determine how accurately we recover the 
intrinsic luminosity function with this prescription for the ``empirical'' 
error functions of the photometric redshifts.  Here the redshift error function
of each galaxy is $p_i(z-z_i;z_i) = \frac{1}{\hat\sigma_z'\,(1+z_i)
\sqrt{2\,\pi}}\,\exp\,[-(z-z')^2/2\,\hat\sigma_z'^2\,(1+z_i)^2]$ with 
$\hat\sigma_z' = 0.15$.  The results are shown in Figure 8 as well.  The solid
curve shows the best-fit Schechter luminosity function.  The best-fit $M_*$ and
$\alpha$ are the lower solid point in the inset with the 99\% uncertainties
indicated by the solid contour.  The step function shows the best-fit stepwise
luminosity function with the vertical bars indicating the associated 
$3\,\sigma$, one-parameter uncertainties.  Comparison of the simulation results
for the luminosity function measurements with the input model shows that we are
able to significantly reduce the systematic uncertainties (by $\sim$ 0.7 mag) 
by taking into account an empirical redshift error function for each galaxy in
the STY and SWML approaches.  Although a lack of data means we could not 
constrain the faint-end galaxy luminosity function very well, we were able to 
recover the bright-end galaxy luminosity function to approximately one 
magnitude fainter than $M_*$.  

\subsection{The Galaxy Luminosity Functions from the LCIR Photometric Redshift
Survey}

  We applied the modified maximum likelihood analysis to determine the 
rest-frame $R$-band galaxy luminosity function for $\sim 3000$ $H$-band 
selected galaxies identified at $0.5\leq z \leq 1.5$ in the LCIR survey.  We 
adopted the redshift error function as formulated in equation (3), which is a
convolution of the redshift likelihood function and a Gaussian kernel of width 
$\hat\sigma_z(1+z)$ with $\hat\sigma_z=0.08$ as determined empirically from 
the comparison of photometric and spectroscopic redshifts described in \S\ 2.3.

  The results of the analysis are presented in Figures 9 and 10 for the 
$H$-band detected galaxies.  We calculated the luminosity functions for the 
total $H$-band selected sample (left panels) and for galaxies that have a 
best-fit spectral template of either E/S0 or Sab in the photometric redshift 
likelihood analysis (right panels).  The solid curves in Figure 9 represent the
best-fit Schechter luminosity functions at $0.5 \leq z \leq 0.75$ (top panels),
$0.75 \leq z \leq 1.0$ (middle panels), and $1.0 \leq z \leq 1.5$ (bottom 
panels) determined with the modified STY approach.  Only galaxies identified in
the HDFS region were included in the luminosity function calculations at $0.5 
\leq z \leq 0.75$ for the reasons discussed in \S\ 2.4.  For comparison, we
include the $r*$-band luminosity function determined for galaxies at $z<0.2$
from the Sloan Digital Sky Survey (SDSS; Blanton 2001; dotted curve) in the 
top-left panel.  In addition, the best-fit luminosity functions of the LCIRs 
galaxies at $0.5 \leq z \leq 0.75$ are also indicated in the middle and bottom 
panels as the short-dashed curves.  Because of the lack of sensitivity to 
intrinsically faint galaxies in the $H$-band survey, the solid curves shown in 
the middle and bottom panels were determined by fixing $\alpha$ to the best-fit
value at $0.5 \leq z \leq 0.75$ and allowing only $M_{R_*}$ to vary in the 
likelihood analysis.  We have, however, attempted to determine both $\alpha$ 
and $M_{R_*}$ for galaxies at $0.75 \leq z \leq 1.0$ and the results are shown 
as the long-dashed curves in the middle panels.  The best-fit galaxy luminosity
functions determined based on the modified SWML approach are shown as solid 
points in Figure 9.  The vertical bars indicate the corresponding $1\,\sigma$ 
one-parameter uncertainties determined according to $2\,\ln\,{\cal L}_\phi= 
2\,\ln\,{\cal L}_{\phi,{\rm max}} - \Delta \chi^2$ with $\Delta \chi^2=1$.

  The good agreement at the bright end between the two best-fit Schechter
luminosity functions (the solid and long-dashed curves) in each of the two 
middle panels of Figure 9 shows that despite the lack of sensitivity at the 
faint end, we can determine the bright-end galaxy luminosity function and 
therefore $M_{R_*}$ fairly reliably using the $H$-band selected LCIR survey 
sample.  In addition, a comparison of the solid and short-dashed curves in
each panel indicates that there is little or no evolution in the rest-frame 
$R$-band galaxy luminosity function for the entire $H$-band selected sample.
The 99\% error contours of the best-fit $M_{R_*}$ and $\alpha$ using the 
modified STY analysis are shown in Figure 10 for galaxies in different redshift
intervals.  It is clear from the top panels that the faint-end slope for 
galaxies at these redshifts is poorly constrained because our survey is 
insensitive to intrinsically faint galaxies at $z\geq 0.75$.

  We summarize the results of the modified STY analysis in Table 1, in which we
list the redshift range, the best-fit faint-end slope $\alpha$, $M_{R_*}-
5\,\log\,h$, and $\phi_*$ together with the associated $1\,\sigma$, 
one-parameter errors.  The number of galaxies that are included in the analysis
for all $H$-band selected galaxies and the $H$-band selected early-type 
galaxies are also listed.  The results of the modified SWML analysis are 
summarized in Table 2, in which we list the rest-frame $R$-band magnitude, $M_R
-5\,\log\,h$, and the best-fit $\log\,\phi (M_R)/(h^3 {\rm Mpc}^{-3})$ together
with the associated $1\,\sigma$, one-parameter errors in three different 
redshift ranges for the $H$-band selected total and early-type samples.

\section{Discussion}

\renewcommand{\thefootnote}{\fnsymbol{footnote}}
\setcounter{footnote}{0}

 We have developed a technique to obtain a robust estimate of the intrinsic 
luminosity function using photometric redshifts.  We adopt an ``empirical'' 
estimate of the redshift error function for each galaxy, which is a convolution
of the redshift likelihood function that characterizes the uncertainty due to
photometric errors and a Gaussian kernel of 1-$\sigma$ width $\sigma_z$ that
characterizes the uncertainty due to template mismatch.  This technique takes 
into account the non-gaussian error characteristics of photometric redshifts, 
and thus differs from the approach of Subbarao \etal\ (1996), who used a simple
Gaussian to model the photometric redshift uncertainties.  The results of the
simulation described in \S\ 3.4 show that our approach allows us to completely
recover the galaxy luminosity function to the sensitivity limit of the survey.
Here we compare the derived galaxy luminosity functions at different redshifts
and discuss the implications for the evolution of galaxies since $z\sim 1.5$.

\subsection{Galaxy Evolution at $0.5\leq z \leq 1.5$}

  It is clear from Figure 9 that the depth of the $H$-band survey is not 
sufficiently sensitive to constrain the statistical properties of galaxies 
fainter than $\sim L_*$ at $z\geq 0.75$, and we are therefore unable to 
determine the evolution of the faint galaxy population based on the $H$-band 
survey.  However, the data are sufficient to determine the evolution of the 
bright galaxy population at $0.5\leq z \leq 1.5$ based on the $H$-band survey 
data.  The evolution of this population has significant implications for 
understanding how galaxies form because different formation scenarios make
distinct predictions for the number density evolution of massive galaxies.  For
example, under monolithic collapse scenarios (e.g.\ Eggen, Lynden-Bell, \& 
Sandage 1962), massive galaxies form early on a dynamical timescale and have a
constant co-moving space density and gradually declining intrinsic 
luminosities.  In contrast, hierarchical formation scenarios (e.g.\ White \& 
Rees 1978) predict that massive galaxies form through the merger of lower-mass
galaxies over a Hubble time and therefore the co-moving space density of 
massive galaxies increases as the universe evolves.

  The luminosity functions presented in Figure 9 and Tables 1 and 2 show 
several interesting results.  First, the best-fit Schechter luminosity 
functions agree with the stepwise luminosity functions to within the
measurement uncertainties at all redshifts, including the luminosity function 
at $0.5\leq z\leq 0.75$ where our survey is sufficiently sensitive to identify
galaxies as faint as $\sim 0.2\,L_*$.  Second, there exists little or no 
evolution either in $M_{R_*}$ or in $\phi_*$ for the total $H$-band selected 
population at $0.5\leq z \leq 1.5$.  Third, we find a moderate luminosity
evolution for galaxies with an SED best characterized as E/S0 or Sab galaxies.
Specifically, our measurements obtained from the modified STY approach suggest 
that from $z\sim 1$ to $z\sim 0.5$ these early-type galaxies have faded by 
$\sim 0.5$ mag.  This is consistent with the expected luminosity evolution for 
a galaxy formed at $z_f=30$ and following an exponentially declining star 
formation rate model with $\tau = 1$ Gyr (McCarthy \etal\ 2001; Chen \etal\ 
2002).  Finally, we find only a mild evolution in $\phi_*$ between $z\approx 
0.5$ and $z \approx 1$ ($\sim 40$\% decrease with redshift) for the 
color-selected early-type galaxies, but we cannot argue against the no 
evolution scenario at more than the $2\,\sigma$ level.  When we consider the 
entire $H$-band selected sample at $0.5\leq z \leq 1.5$, our estimates of 
$\phi_*$ are completely consistent with a no evolution scenario.  

  It appears that neither the total $H$-band selected sample nor the 
color-selected early-type galaxies have evolved significantly in intrinsic 
luminosity or space density since $z\sim 1.5$, contrary to the predictions of 
the hierarchical formation scenarios.  But as with all luminosity function
measurements, it is clear that there exists a strong degeneracy between 
$M_{R_*}$ and $\phi_*$ at $z\geq 0.75$ in our analysis, because the $H$-band 
survey is not sensitive to sub-$L_*$ galaxies at this redshift range and our 
estimates for these two parameters were determined from the space density of 
intrinsically bright galaxies with a fixed faint-end slope $\alpha$.  For 
example, a bright $M_{R_*}$ together with a small $\phi_*$ can produce the same
bright-end luminosity function as a faint $M_{R_*}$ with a large $\phi_*$.  To
estimate how significantly early-type galaxies have evolved since $z\sim 1.5$,
we therefore calculated the rest-frame co-moving $R$-band luminosity density 
$\ell_R$ for galaxies brighter than $M_{\rm min}$ using the best-fit Schechter 
luminosity function, 
\begin{eqnarray}
\ell_R &=& \int_{L_{min}}^\infty L\cdot \Phi(L;L_*)\,d(L/L_*) \\
       &=& \int_{-\infty}^{M_{\rm min}} (0.4 \ln 10)\cdot\phi_*\cdot L_*\cdot 10^{0.4\,(M_*-M)(2+\alpha)}\cdot\exp(-10^{0.4\,(M_*-M)})\,dM.
\end{eqnarray}
This represents an integrated quantity that characterizes the brightest galaxy
population and is least sensitive to faint galaxies.  

  We test different formation scenarios using the estimated co-moving 
luminosity density evolution of the bright, red galaxies.  Including the 
luminosity function estimated for early-type galaxies at $z=0.3$ in the CNOC2 
survey (Lin \etal\ 1999), we first compared $\ell_R$ calculated for a constant 
$M_{\rm min}$ at all redshifts.  The results are presented in the left panel of
Figure 11, which shows a flat $\ell_R$ versus $z$ for $M_{\rm min} = -20.5$ 
(squares) , $-21.0$ (triangles), and $-21.5$ (circles).  These roughly 
corresponds to $L_*$, $1.6\,L_*$, and $2.5\,L_*$, respectively, at $z\sim 0.3$.
However, because stars evolve with time, the corresponding luminosity evolution
must be accounted for.  To distinguish between a passively evolving population
and hierarchical mergers, we estimated the amount of brightening ($\Delta\,
M_R$) with increasing redshift (or with decreasing age of the universe) for 
different stellar evolution models (see Table 3 for a summary).   We considered
a wide range of stellar evolution scenarios, from a single burst at $z_f=30$, 
to an exponentially declining star formation rate model with $z_f=5$ and $\tau
= 1$ Gyr.  The former represents a classic passive evolution mode under the 
monolithic collapse scenariol.  The latter represents the closest resemblance 
of a more continuous star-forming scenario in which some residual star 
formation is taking place at $z<3$, while there is still enough time for the 
red galaxy population to be present by $z\sim 1$.  We then calculated $\ell_R$
for a suite of evolving $M_{\rm min}(z)$ according to different stellar 
evolution recipes.  

  The goal of this analysis is to examine whether or not mergers are the 
dominant process for the formation of red galaxies.  Mergers would lead to 
brightening with time (smaller systems combined to form bigger ones), which 
compensates the expected fading due to stellar evolution.  We would therefore 
expect to observe an increasing $\ell_R$ with decreasing redshift using 
($\Delta\,M_R$) presented in Table 3, if these red galaxies formed through 
a sequence of mergers.  A flat redshift distribution of $\ell_R$ would then 
imply that mergers are insignificant in the redshift interval tested.

  We present the results in Figure 11.  The middle panel is for the $z_f=30$
single burst scenario and the right panel is for the $z_f=5$ exponentially 
declining star formation rate model.  The results in these two panels have been
scaled to have consistent $M_{\rm min}$ at $z=0.3$ as in the left panel.  In 
addition, we exclude the brightest subsample (circles in the left panel) from 
these two panels, because the number of galaxies at the very bright end of the
luminosity functions is very small beyond $z\sim 0.75$.  Specifically, there 
are only eight galaxies in the red sample that are brighter than $M_R=-23$ at 
$1.0 \leq z \leq 1.5$.

  Figure 11 shows that $\ell_R$ evolves relatively slowly with redshift for
different bright subsamples, and that it could have increased by at most a 
factor of $\approx 6$ from $z\sim 1.2$ to $z\sim 0.3$ for the brightest 
early-type sample under the most extreme stellar evolution model presented in 
the right panel.  Taking into account the fact that these models do not have 
dust included, which may further reduce the amount of fading in intrinsic 
luminosity with redshift, we find no evidence to support that mergers dominate 
the formation of early-type galaxies over $0.5 \leq z \leq 1.5$.  This is 
consistent with the findings of Firth \etal\ (2002), who directly compared the 
observed abundances of red galaxies at $z\sim 1$ with those from different 
semi-analytical models.  They found that the models underpredict the number of 
red galaxies observed in the LCIR survey by at least a factor of three.

  The rest-frame co-moving $R$-band luminosity densities for all $H$-band
selected galaxies at different redshifts are presented in Figure 12.  As in
Figure 11, we calculated $\ell_R$ for three constant thresholds, $M_{\rm min}=
-20.5$ (squares) , $-21.0$ (triangles), and $-21.5$ (circles).  Because 
galaxies in the total $H$-band selected sample have a wide range of star 
formation history, we cannot remove the effects of pure luminosity evolution by
adopting a simple $M_{\rm min}(z)$.   The results of Figure 12 show that there
exists only a mild evolution in the rest-frame co-moving $R$-band luminosity 
density for the entire galaxy population over $0.5 \leq z\leq 1.5$.

  It is also important to note that our analysis is based on an $H$-band 
selected galaxy sample, which roughly corresponds to galaxies selected in the 
rest-frame $J$, $I$, and $R$ bands at $z\sim 0.5$, 0.9, and 1.3, respectively. 
A potential systematic bias in our measurements is clearly the bandpass 
selection effect which, when combined with variation in star formation rate
and stellar evolution with time, makes it difficult to interpret our results, 
particularly for the total $H$-band selected sample.  At $z\sim 1.3$, the 
$H$-band survey preferentially selects optically bright (and therefore younger)
galaxies, while it tends to select near-infrared bright (and therefore older) 
galaxies at $z\sim 0.5$, when the universe was twice as old.  Because the total
$H$-band selected sample consists of galaxies of different star formation
histories in different redshift intervals, it is possible that galaxies of
different colors, selected at different rest-frame wavelengths at different
redshifts conspire to produce a luminosity function that appears to show 
little evolution.  Consequently, it is impossible to rule out any evolution 
scenarios without a detailed understanding of the intrinsic properties of the
galaxies in the total $H$-band sample.  On the other hand, the bandpass 
selection bias has a much less serious effect for the early-type galaxy sample.
Only galaxies with an observed SED (spanning a spectral range from the $V$ band
or $U$ for the HDFS objects through the $H$ band) that is consistent or redder 
than a typical Sab galaxy were included in the analysis at all redshifts.  Our 
measurements represent a true, empirical measure of how ``red'' galaxies 
evolve with time, and the results of our analysis suggest that early-type 
galaxies have not evolved significantly since $z\sim 1.5$.

\subsection{Comparison with Previous Studies}

  The luminosity functions presented in \S\ 3.6 cover the redshift range $0.5 
\leq z \leq 1.5$.  The upper end of this range extends beyond the reach of
traditional spectroscopic redshift surveys, the most ambitious of which stretch
to $z=1.2$ but become significantly incomplete beyond $z=1$ (Lilly \etal\ 1995;
Cowie \etal\ 1996; Cohen 2002).  Various groups have measured evolution in the 
luminosity function using spectroscopic redshifts over the redshift range $0 < 
z < 1$ for galaxies of different colors (e.g.\ Lilly \etal\ 1995), different 
spectral types (e.g.\ Heyl \etal\ 1997; Cohen 2002), or different morphology 
(Im \etal\ 1999).  In addition, some groups have attempted to measure the 
luminosity function using photometric redshifts for galaxies at $z\apl 4$ (Gwyn
\& Hartwick 1996; Mobasher \etal\ 1996; Connolly \etal\ 1997; Sawicki \etal\ 
1997).  Different sample selection criteria make a quantitative comparison 
between these measurements difficult.  Here we present a qualitative comparison
between our measurements and the previous ones.

  First, we consider previous measurements that were conducted in the 
rest-frame $R$ band.  The {\it local} luminosity density is now well 
constrained by both the 2dF Redshift Survey (Cross \etal\ 2001) and the SDSS 
(Blanton 2001).  It is encouraging that these surveys yielded consistent 
results in the $B$ band.  The $R$-band luminosity density available from the 
SDSS is most directly relevant to our results from the $H$-selected LCIR 
survey, and is included in Figure 12 for the luminosity density evolution of 
the total $H$-band selected sample.  If we consider only the LCIR measurements 
at $0.5 < z < 1$ and the SDSS measurement at $z < 0.2$, then the evolution in 
the luminosity density from $z \sim 0$ to $z\sim 1$ is less than 0.2 dex for 
any reasonable choice of the faint-end slope, $\alpha$, indicating a very slow 
evolution.

  Next, we compare our results with those from a number of very deep redshift 
surveys that covered a relatively smaller area but were sensitive to galaxies 
at redshift beyond $z\sim 1$.  These are the Canada-France Redshift Survey 
(CFRS; e.g.\ Lilly \etal\ 1996), a deep $K$-band survey by Cowie, Songaila, \& 
Barger (1999), and the Caltech Faint Galaxy Redshift Survey (CFGRS; Cohen 
\etal\ 2002), all of which showed that the bulk of the evolution in the 
luminosity function occured in bluer spectral types.  In addition, they showed
that galaxies with redder SEDs evolved more slowly, if at all, over this 
redshift range, a result with which the LCIR survey is at least qualitatively 
consistent.  Note however that these results may change if adopting a different
color selection criterion for red galaxies at high redshifts (Kauffmann, 
Charlot, \& White 1996) or the incompleteness correction for absorption feature
dominated galaxies was incorrect (Totani \& Yoshii 1997), two effects that work
in opposite ways.

  A more quantitative comparison is possible with Lilly \etal\ (1996), which
gave the evolution of the luminosity density between $z=0$ and $z=1$ for 
different CFRS samples.  In particular, Figure 12 is most directly comparable 
to Figure 1 in Lilly \etal\ (1996).  The CFRS luminosity densities were 
computed at rest-frame 2800\,\AA, 4400\,\AA, and 1\,$\mu$m by interpolating 
their $BVIK$ photometry, which showed a decline by approximately 0.5 dex at 
4400\,\AA\ and 0.4 dex at $1\,\mu$m from $z\sim 1$ to $z\sim 0$.\footnote{We 
have subtracted 0.2 dex from the original measurements in Lilly \etal\ to 
account for the differences in the adopted cosmological models.}  This gives 
$\Delta\log \ell\,/\Delta\log (1+z) = 1.7 \pm 0.5$ at 4400\,\AA\ and $1.3 \pm 
0.4$ at $1\,\mu$m.  We find from the LCIR survey that $\Delta\log \ell\,/\Delta
\log (1+z)=0.6 \pm 0.1$ at 6800\,\AA, if we assume a faint-end slope $\alpha=
-1.0$.  If we assume a very steep faint-end slope $\alpha=-1.5$, then we 
measure still only $1.0 \pm 0.2$.  Although our results are marginally 
consistent with the CFRS results, the $H$-selected LCIR survey appears to favor
slower evolution in the overall galaxy population at $z < 1$ (at the $\approx
\,1\,\sigma$ level).  

  The slow evolution measured here is similar to the $I$-band evolution 
measured by Cowie, Songaila, \& Barger (1999) from $K$-selected spectroscopic 
surveys at $z < 1.2$.  Based on a much larger sample covering a broader 
redshift range, we find that this slower evolution at red wavelengths extends 
to at least $z=1.5$.  Our results are also consistent with the interpretation 
of the luminosity function evolution derived from the CFGRS, which represents 
by far the largest and most complete spectroscopic sample of galaxies at $z > 
0.8$.  It is, however, still limited by small sample size (144 galaxies at $0.8
\leq z \leq 1.05$ and 18 galaxies at $1.05 \leq z \leq 1.5$).  This illustrates
the difficulties of carrying out a complete spectroscopic survey over a large 
area to very faint magnitude limits.

  Next, we compare our results with those estimated based on photometric
redshifts.  At $1 < z < 2$, photometric redshift surveys prevail but have so 
far been limited to small areas of the sky.  In addition, it is important to
include near-infrared photometry in order for accurate photometric redshift 
measurements to be plausible.  Connolly \etal\ (1997) showed that uncertainties
of photometric redshifts are reduced by 40\%, when incorporating near-infrared 
photometry in the photometric redshift analysis.  Indeed, the reddest galaxies 
known in this reshift interval (Elston \etal\ 1988; McCarthy \etal\ 1992; Hu \&
Ridgeway 1994) are often exceedingly faint in all optical bandpasses bluer than
the $R$ band, making photometric redshift measurements very uncertain.  The 
luminosity function from Connolly \etal\ was for a $J$-band selected galaxy 
sample over $1 < z < 2$ in the Hubble Deep Field North (HDFN).  Their 
luminosity function was computed, however, at rest-frame 2800 \AA, which is 
more sensitive to young stellar populations than the rest-frame $R$ band.  
Furthermore, over the redshift range probed by the LCIR survey ($0.5 < z < 
1.5$), the HDFN is primarily sensitive to the faint end of the luminosity 
function, while the LCIR survey is sensitive only to the bright end.  Finally,
the total survey area of HDFN was only 4.48 arcmin$^2$, corresponding to a 
comoving volume of $\approx 1.3\times 10^4$ Mpc$^3$ over the $1 \leq z \leq 2$ 
interval.  In contrast, our analysis is based on a survey volume that is more 
than two orders of magnitude larger than in the HDFN.  Therefore, we have 
sufficient numbers of bright galaxies to $z=1.5$ to probe their evolution in 
greater detail.  The larger volume is necessary to accurately measure the mean
space density because the bright, red galaxies are now known to be strongly 
clustered (Daddi \etal\ 2000; McCarthy \etal\ 2001; Firth \etal\ 2002).

  Other groups have attempted to study the evolution of the galaxy luminosity 
function at $z\leq 1$ based on a deep photometric redshift survey that 
incorporates near-infrared photometry (Fried \etal\ 2001; Poli \etal\ 2001), 
but these galaxy samples were selected in the $I$ band, which lends a bluer 
tint to the overall mix of SEDs than the $H$-selected LCIR survey galaxies.  It
is therefore not straightforward to compare our results with these other 
photometric redshift surveys.

  Finally, we consider measurements obtained for early-type galaxies selected 
on the basis of quantitative morphological criteria such as smoothness, 
symmetry, and bulge-to-total ratio.  Im \etal\ (1999) selected a sample of 145 
E/S0 galaxies from deep HST imaging in the Groth Strip (Groth \etal\ 1994)
Using a subset with spectroscopic redshifts as calibrators, they estimated 
redshifts for the remaining galaxies using the $V-I$ color alone.  The 
resulting luminosity functions (measured in the rest-frame $B$ band) suggest 
slow evolution in the E/S0 population at $z < 1$, in qualitative agreement with
the results discussed earlier.  Brinchmann \etal\ (1998) and Menanteau \etal\ 
(1999) have, however,suggested that the bulk of morphologically selected 
early-type galaxies at $z < 1$ have bluer colors than those predicted by 
different stellar evolution models, but the results are inconclusive because 
they depend sensitively on the adopted models and how photometric uncertainties
are incorporated in these models.

  In summary, we find that the rest-frame, co-moving optical luminosity density
of the entire galaxy population has evolved only moderately (declining by a 
factor of $\approx 2$) from $z\sim 1$ to $z\sim 0$.  The rest-frame, co-moving
optical luminosity density of the color-selected, early-type galaxies has not 
evolved significantly since $z\sim 1$.  Our new measurements of luminosity 
density evolution is based on a significantly larger sample selected in the
near-infrared than previous surveys selected at a range of wavelengths.  It
is consistent with previous findings, but we place a much tighter constraint on
the apparant lack of evolution in the co-moving luminosity density of 
early-type galaxies. 

\section{Summary}

  We have obtained an estimate of the galaxy luminosity function at $z \geq 1$
using photometric redshifts for galaxies identified in the HDFS and CDFS
regions of the LCIR survey.  We have also calculated the luminosity functions 
at redshifts between $z\approx 0.5$ and $z\approx 1.0$ using the same galaxy
sample selected uniformly at near-infrared wavelengths.  The luminosity 
function analysis is based on $\sim 3000$ $H$-band selected galaxies with
apparent magnitude $17\leq H\leq 20$ at $0.5\leq z\leq 1.5$.  We demonstrate 
that our photometric redshift measurements are accurate with an RMS dispersion 
between the photometric and spectroscopic redshifts of $\sigma_z/(1+z) \approx 
0.08$.  In addition, we show that the systematic uncertainty inherent in the 
luminosity function measurements due to uncertainties in photometric redshifts 
is non-negligible and therefore must be accounted for.  We develop a technique
to incorporate photometric redshift error functions of individual galaxies in 
the estimates of the intrinsic luminosity function.  We show that this 
technique allows us to completely recover the galaxy luminosity function to the
sensitivity limit of the survey and therefore to study the luminosity and space
density evolution of near-infrared selected galaxies over a large redshift
interval.  The results of our study are:
 
  1. On the basis of galaxies identified in the Hubble Deep Field South and 
Chandra Deep Field South regions, we find that ($7.3\pm 0.2$)\,\% and ($16.7\pm
0.4$)\,\%, respectively, of the $H$-band detected galaxies of $H\leq 20$ are 
at $z\geq 1$.  

  2. The galaxy luminosity functions determined via a modified SWML method are
consistent with the Schechter functions determined via the modified STY method.

  3. The best-fit $M_{R_*}$ and $\phi_*$ for the total $H$-band selected 
population are consistent with a no-evolution model at $0.5\leq z \leq 1.5$.

  4. The evolution of the co-moving luminosity density of the $H$-band selected
galaxies is characterized by $\Delta\log \ell\,/\Delta\log (1+z)=0.7 \pm 0.2$ 
at rest-frame 6800\,\AA.  This is marginally consistent with the CFRS results,
but favors a slower evolution (at the $\approx\,1\,\sigma$ level).

  5. Galaxies with spectral energy distributions best characterized as E/S0 or
Sab appear to have faded by $\approx 0.5$ mag from $z\sim 1$ to $z\sim 0.5$, 
consistent with the expected luminosity evolution from an exponentially 
declining star formation rate model.  In addition, the space density of the 
these color-selected early-type galaxies has evolved only mildly since $z
\approx 1$ ($\sim 40$\% decrease with redshift).

  6. The derived rest-frame co-moving $R$-band luminosity density $\ell_R$ of 
color-selected early-type galaxies exhibits only moderate evolution with
redshift.  Specifically, we find that under the most extreme stellar evolution
scenario the $\ell_R$ of early-type galaxies brighter than $L_*$ ($1.6\,L_*$) 
could have increased by {\em at most} a factor of $\approx 3 (6)$ from $z\sim 
1.2$ to $z\sim 0.3$.  Our results place an upper limit to the merger rate of 
bright galaxies over this time interval.

\acknowledgements

  We appreciate the expert assistance from the staffs of the Cerro Tololo 
Inter-American Observatory and the Las Campanas Observatory. We thank Andrew
Firth, Chris Sabbey, Richard McMahon, and Kathleen Koviak for assistance with 
the observations.  This research was supported by the National Science 
Foundation under grant AST-9900806.  The CIRSI camera was made possible by the
generous support of the Raymond and Beverly Sackler Foundation.

\newpage

\begin{deluxetable}{ccccrrcccr}
\tiny
\tablecaption{Results of the modified STY Analysis\tablenotemark{a}}
\rotate
\tablewidth{0pt}
\tablehead{\colhead{} & \multicolumn{4}{c}{All} & \colhead{} & 
\multicolumn{4}{c}{E/S0 $+$ Sab} \\
\cline{2-5} 
\cline{7-10} \\
\colhead{} & \colhead{} & \colhead{$M_{R_*}$} & \colhead{$\phi_*$} & \colhead{
Number} & \colhead{} & \colhead{} & \colhead{$M_{R_*}$} & \colhead{$\phi_*$} & 
\colhead{Number} \\
\multicolumn{1}{c}{Redshift Range} & \colhead{$\alpha$} & 
\colhead{$-5\,\log\,h$} & \colhead{$(h^3 {\rm Mpc}^{-3})$} & 
\colhead{of Galaxies} & \colhead{} & \colhead{$\alpha$} & \colhead{$-5\,\log\,h$} & \colhead{$(h^3 {\rm Mpc}^{-3})$} & \colhead{of Galaxies}}
\startdata
 $[0.50,0.75]$ & $-1.00_{-0.02}^{+0.06}$ & $-21.03_{-0.10}^{+0.03}$ & $0.0142_{-0.0013}^{+0.0012}$ & 978 & & $-0.20_{-0.04}^{+0.05}$ & $-20.43_{-0.03}^{+0.01}$ & $0.0127_{-0.0014}^{+0.0015}$ & 606 \nl
 $[0.75,1.00]$ & $-0.50_{-0.06}^{+0.08}$ & $-21.03_{-0.05}^{+0.04}$ & $0.0142_{-0.0013}^{+0.0015}$ & 1239 & & $+0.20_{-0.11}^{+0.09}$ & $-20.83_{-0.04}^{+0.05}$ & $0.0070_{-0.0009}^{+0.0009}$ & 752 \nl
                         & $-1.00_{-0.00}^{+0.00}$ & $-21.44_{-0.07}^{+0.07}$ & $0.0103_{-0.0010}^{+0.0011}$ & 1239 & & $-0.20_{-0.00}^{+0.00}$ & $-21.04_{-0.06}^{+0.08}$ & $0.0072_{-0.0009}^{+0.0009}$ & 752 \nl
 $[1.00,1.50]$ & $-1.00_{-0.00}^{+0.00}$ & $-20.92_{-0.03}^{+0.02}$ & $0.0271_{-0.0039}^{+0.0047}$ & 791 & & $-0.20_{-0.00}^{+0.00}$ & $-20.67_{-0.03}^{+0.04}$ & $0.0111_{-0.0020}^{+0.0024}$ & 458 \nl
\tablenotetext{a}{Numbers with zero errors indicate that the parameters were 
fixed at the listed values in the likelihood analysis.}
\enddata
\end{deluxetable}


\begin{deluxetable}{ccccrccc}
\tablecaption{Modified SWML Galaxy Luminosity Function, $\log\,\phi(M_R)/(h^3 {\rm Mpc}^{-3})$}
\tablewidth{0pt}
\tablehead{\colhead{} & \multicolumn{3}{c}{All} & \colhead{} & 
\multicolumn{3}{c}{E/S0 $+$ Sab} \\
\cline{2-4} 
\cline{6-8} \\
\multicolumn{1}{c}{$M_R-5\,\log\,h$} & \colhead{$[0.50,0.75]$} & 
\colhead{$[0.75,1.00]$} & \colhead{$[1.00,1.50]$} & \colhead{} & 
\colhead{$[0.50,0.75]$} & \colhead{$[0.75,1.00]$} & \colhead{$[1.00,1.50]$}}
\startdata
$-23.23$ & ... & $-7.15_{-0.16}^{+1.42}$ & ... & & ... & $-4.57_{-0.22}^{+0.22}$ & ... \nl
$-22.73$ & $-5.52_{-0.01}^{+1.07}$ & $-5.01_{-0.58}^{+0.65}$ & $-4.63_{-0.04}^{+0.77}$ & & ... & $-4.75_{-0.97}^{+0.74}$ & $-4.92_{-0.15}^{+0.83}$ \nl
$-22.23$ & $-3.73_{-0.10}^{+0.32}$ & $-3.20_{-0.02}^{+0.28}$ & $-2.99_{-...}^{+0.02}$ & & $-5.23_{-0.01}^{+1.20}$ & $-2.88_{-0.14}^{+0.03}$ & $-3.46_{-0.01}^{+0.01}$ \nl
$-21.73$ & $-2.96_{-0.06}^{+0.18}$ & $-2.27_{-0.09}^{+0.01}$ & $-2.54_{-0.01}^{+0.04}$ & & $-3.70_{-0.04}^{+0.48}$ & $-2.64_{-0.05}^{+0.04}$ & $-2.60_{-0.03}^{+0.04}$ \nl
$-21.23$ & $-2.27_{-0.03}^{+0.09}$ & $-2.21_{-0.02}^{+0.04}$ & ... & & $-2.65_{-0.02}^{+0.35}$ & $-2.71_{-0.04}^{+0.04}$ & ... \nl
$-20.73$ & $-1.93_{-0.06}^{+0.02}$ & $-2.15_{-0.05}^{+0.03}$ &...  & & $-2.19_{-0.16}^{+0.01}$ & $-2.25_{-0.04}^{+0.06}$ & ... \nl
$-20.23$ & $-1.94_{-0.02}^{+0.02}$ & ... & ... & & $-2.18_{-0.01}^{+0.03}$ & ... & ... \nl
$-19.73$ & $-2.21_{-0.06}^{+0.03}$ & ... & ... & & $-2.58_{-0.07}^{+0.03}$ & ... & ... \nl
\enddata
\end{deluxetable}

\begin{deluxetable}{p{2.25in}rcccc}
\tablecaption{Predicted Brightening ($\Delta\,M_R$) with Redshift for Different
Star Formation Histories}
\tablewidth{0pt}
\tablehead{\colhead{Scenario} & \multicolumn{1}{c}{$z_f$} & \colhead{$z=0.3$} 
& \colhead{$z=0.625$} & \colhead{$z=0.875$} & \colhead{$z=1.25$}}
\startdata
single burst with $\Delta\,t=1$ Gyr \dotfill & 30 & $-0.30$ & $-0.53$ 
	& $-0.69$ & $-0.94$ \nl
             & 10 & $-0.32$ & $-0.55$ & $-0.74$ & $-0.98$ \nl
$\exp(-t/\tau)$ with $\tau=1$ Gyr   \dotfill & 30 & $-0.30$ & $-0.60$ 
	& $-0.81$ & $-1.15$ \nl
             & 10 & $-0.34$ & $-0.64$ & $-0.88$ & $-1.24$ \nl
             &  5 & $-0.35$ & $-0.70$ & $-0.96$ & $-1.40$ \nl
\enddata
\end{deluxetable}

\newpage

\begin{figure}
\plotone{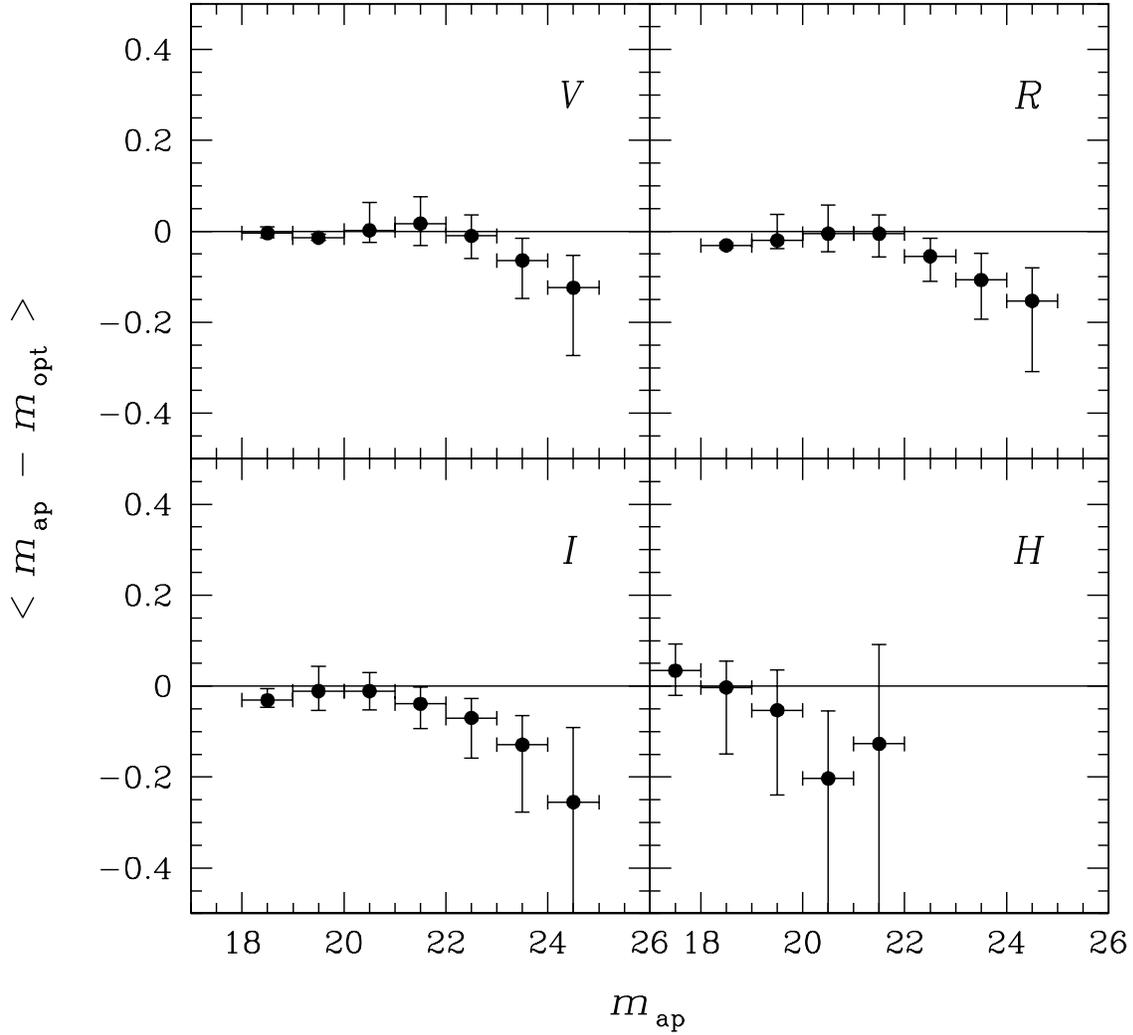}
\caption{Median residuals of ``optimal'' ($m_{\rm opt}$) vs. 4''-aperture 
photometry ($m_{\rm ap}$) in the $V$, $R$, $I$, and $H$ bandpasses for the 
$H$-band selected objects in the CDFS region.  Horizontal bars show the 
magnitude bin size.  Vertical bars mark the $1\,\sigma$ scatters of the 
residuals for individual magnitude bins.  Straight line segment shows $m_{\rm 
opt} = m_{\rm ap}$.}
\end{figure}

\newpage

\begin{figure}
\plotone{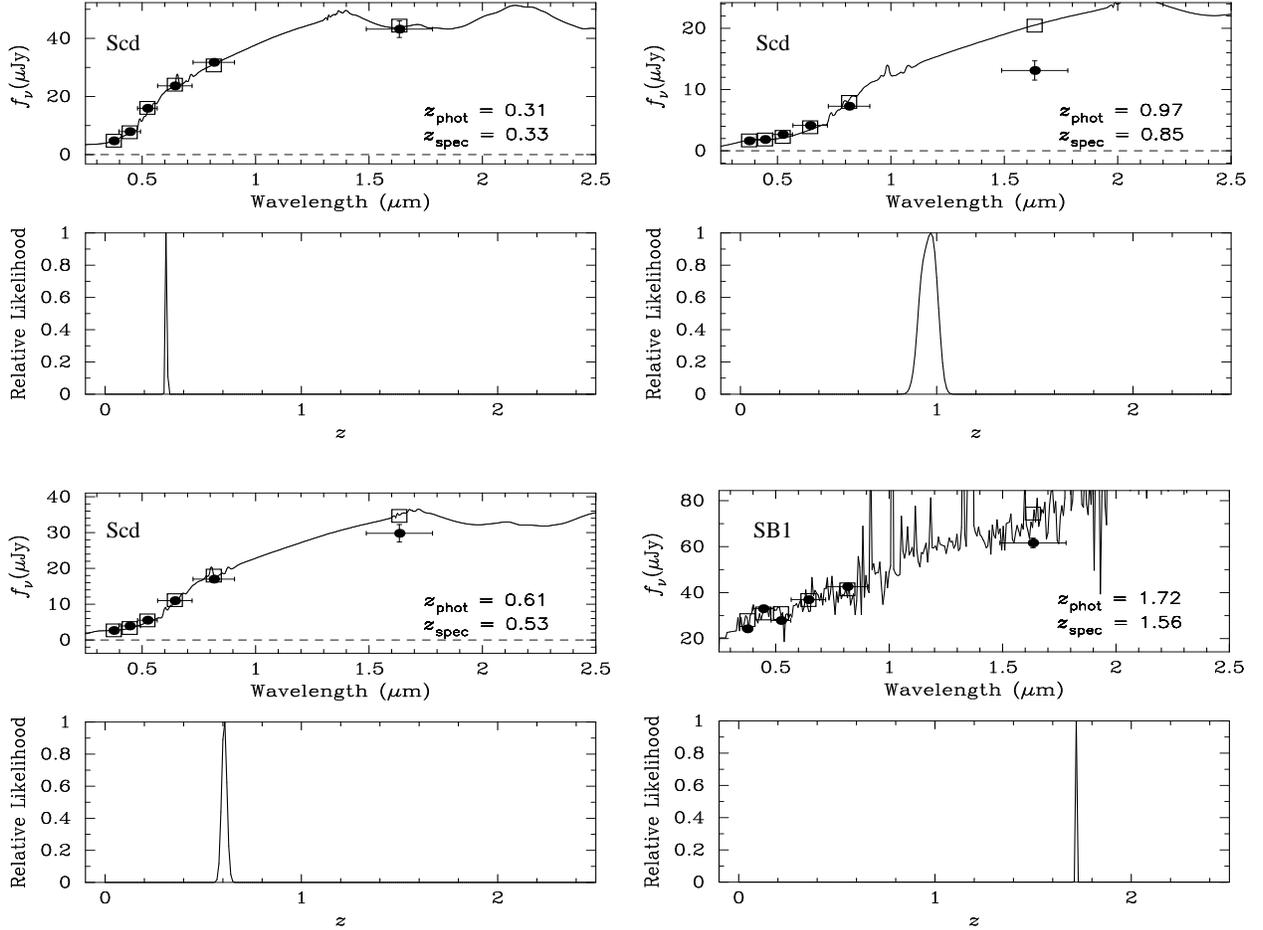}
\caption{Sample results of the redshift likelihood analysis for four 
spectroscopically identified galaxies in the HDFS region.  The top panel of 
each pair shows the photometric measurements and uncertainties as solid points
with error bars.  The solid curves represent the best-fit spectral templates 
with the spectral type indicated in the upper-left corner.  The open squares 
indicate the predicted fluxes in individual bandpasses, estimated by 
integrating the best-fit template over the corresponding system transmission
functions.  The dashed lines indicate the zero flux level.  The bottom panel 
of each pair shows the redshift likelihood functions.  The estimated 
photometric redshift ($z_{\rm phot}$) and spectroscopic redshift ($z_{\rm 
spec}$) are indicated in the lower-right corner of each of the top panels.}
\end{figure}

\newpage

\begin{figure}
\plottwo{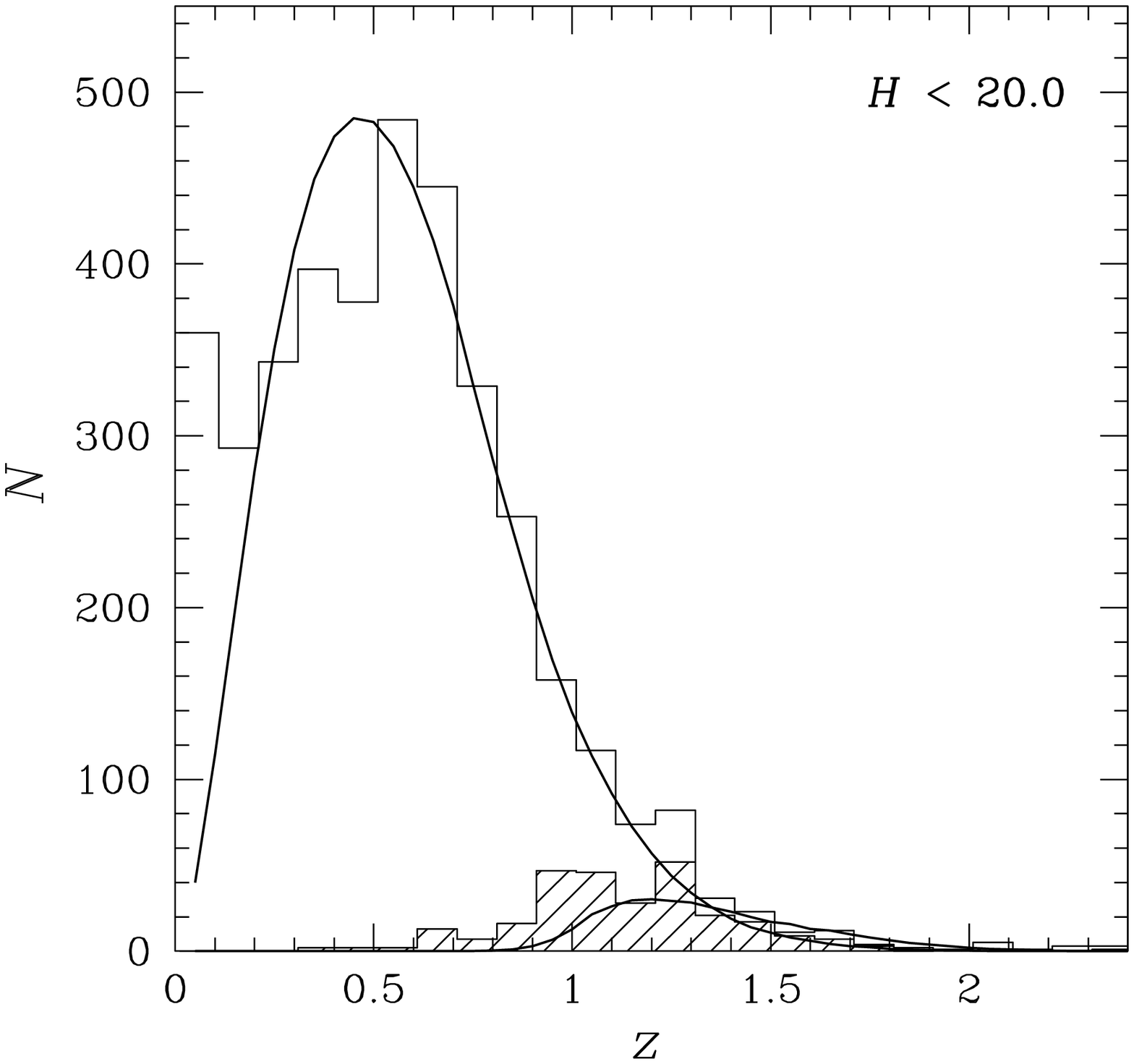}{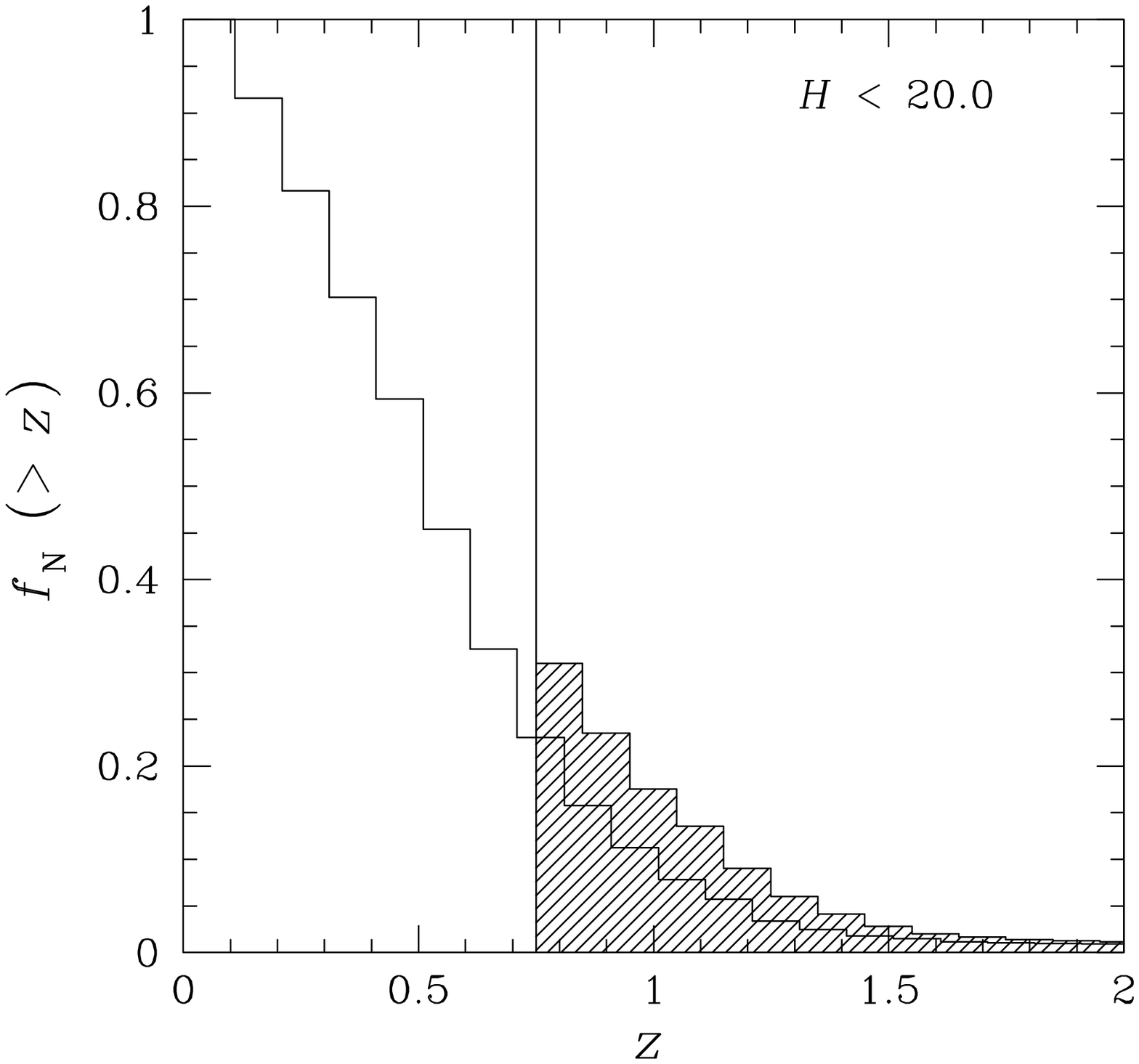}
\caption{Left: Redshift histogram of the $H$-band detected galaxies in the HDFS
region.  The hatched region indicates the redshift distribution of the $I-H\geq
3$ galaxies.  Solid curves indicate the predicted redshift distribution based 
on the model that best fits the number--magnitude relation presented in Chen
\etal\ (2002).  Right: Cumulative redshift distributions of the fraction of
the $H$-band selected galaxies with $H < 20$.  The open and hatched histograms 
indicate galaxies identified in the HDFS and CDFS regions, respectively.  We 
find that between $\approx$ 7\% and 17\% of the $H$-band selected galaxies are
at $z\geq 1$.}
\end{figure}

\newpage

\begin{figure}
\plottwo{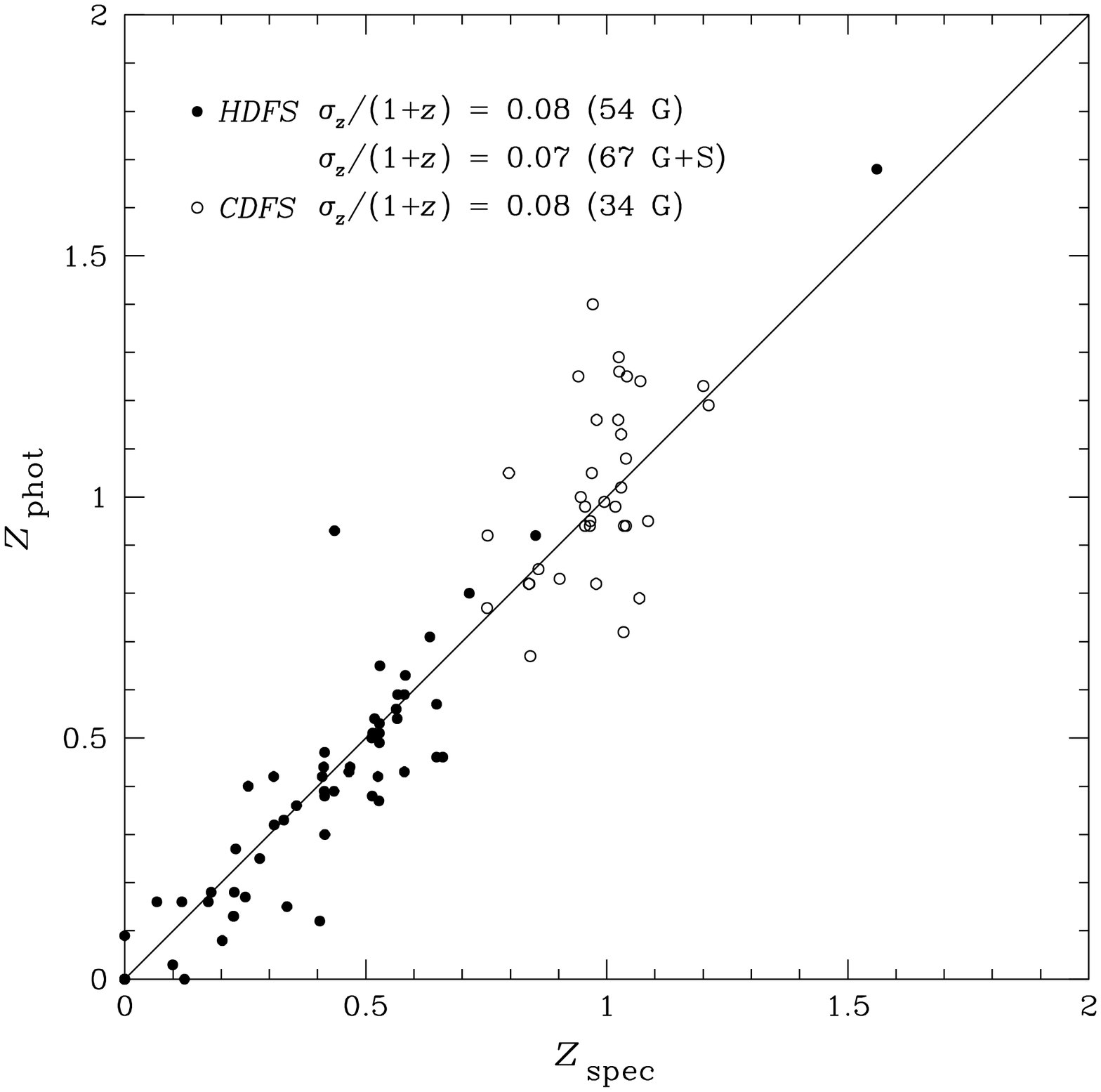}{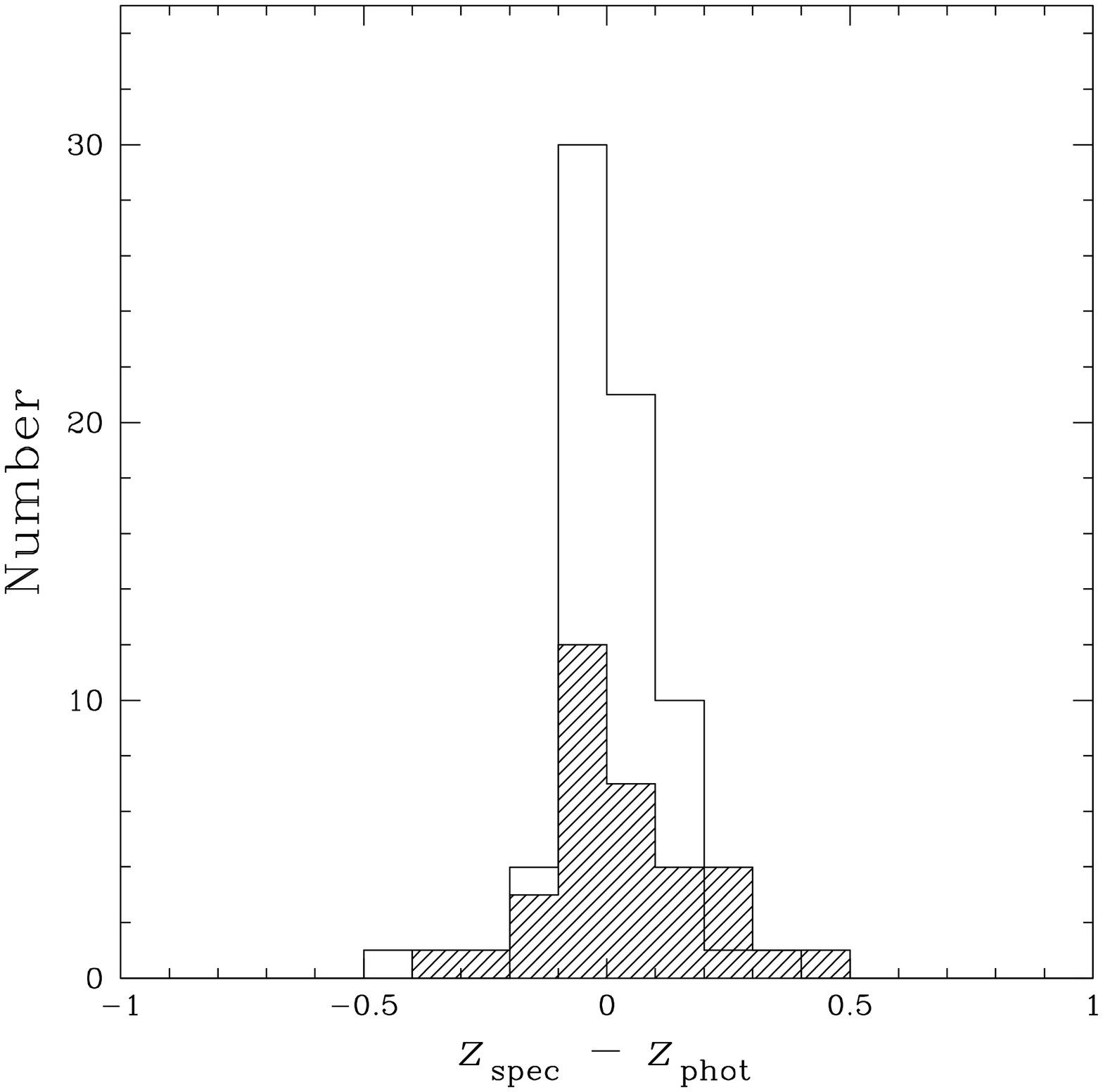}
\caption{Left: Comparisons of photometric ($z_{\rm phot}$) and spectroscopic 
($z_{\rm spec}$) redshifts for the HDFS (solid points) and CDFS (open circles)
objects.  Right: The corresponding redshift residual distributions for the HDFS
(open histogram) and CDFS (shaded histogram) objects.  The solid line in the 
left panel indicates $z_{\rm phot}=z_{\rm spec}$.  There are 67 objects in the 
HDFS region with known spectroscopic redshifts.  We find that photometric 
redshifts are accurate to within $\sigma_z/(1+z) \approx 0.08$ at $z\leq 1$.  
Based on 34 CDFS objects with reliable photometric and spectroscopic redshifts,
we find that photometric redshifts estimated without $U$ and $B$ photometry are
accurate to within $\sigma_z/(1+z) \approx 0.08$ at $z_{\rm spec} > 0.75$.} 
\end{figure}

\newpage

\begin{figure}
\epsscale{0.95}
\plotone{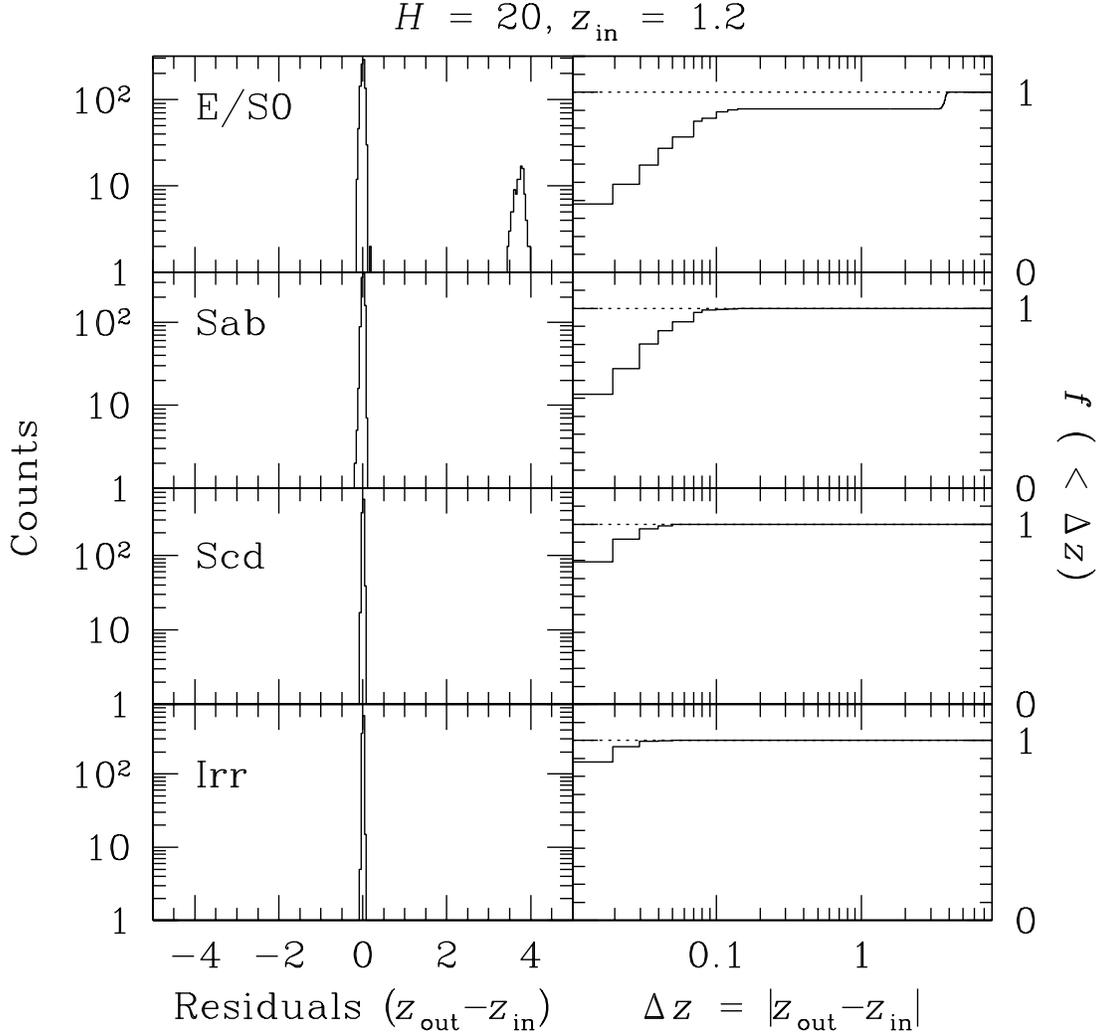}
\caption{Sample results of the Monte Carlo simulations to determine photometric
redshift uncertainties for four different spectral types due to uncertainties 
in the galaxy photometry.  Based on 1000 realizations of a galaxy of $H=20$ at 
$z=1.2$ in the HDFS region, the left panels show histograms of the residuals 
between the input galaxy redshift $z_{\rm in}$ and the best-fit photometric 
redshift $z_{\rm out}$, and the right panels show cumulative fractions of the
simulated photometric redshifts versus absolute redshift residuals.  In each 
realization, we perturbed the photometric measurements within the $1\,\sigma$
photometric uncertainties known for the individual optical and $H$ images in 
the HDFS region.  We find that for a galaxy of $H=20$ at $z=1.2$ in the HDFS 
region there is an $\approx 10\,$\% chance that it would be misidentified as a
starburst galaxy at $z\sim 4$, if the spectral shape mimicked an E/S0 galaxy.}
\end{figure}

\newpage

\begin{figure}
\epsscale{1.05}
\plotone{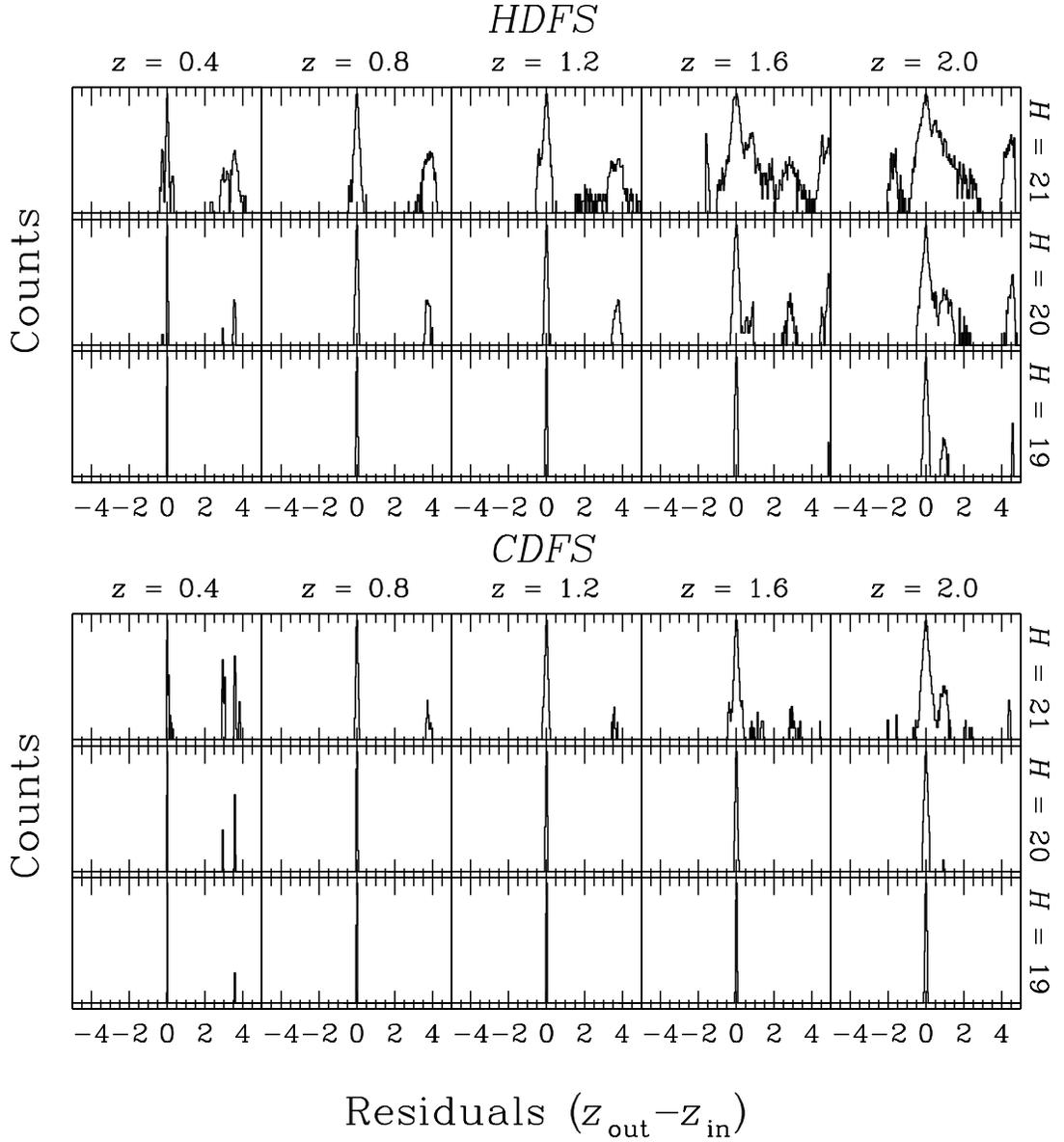}
\caption{Summaries of the Monte Carlo simulations to determine photometric
redshift uncertainties due to uncertainties of galaxy photometry versus
redshift and $H$-band magnitude.  For a given input redshift and $H$-band 
magnitude, each panel now shows the sum of the residual histograms over four 
spectral types as shown separately in Figure 5.  The vertical axis is on a 
logarithmic scale as shown in Figure 5.  The upper panels are results based on
the filter combination and image sensitivities of the HDFS galaxies and the 
lower panels for the CDFS galaxies.}
\end{figure}

\newpage

\begin{figure}
\epsscale{1.05}
\plotone{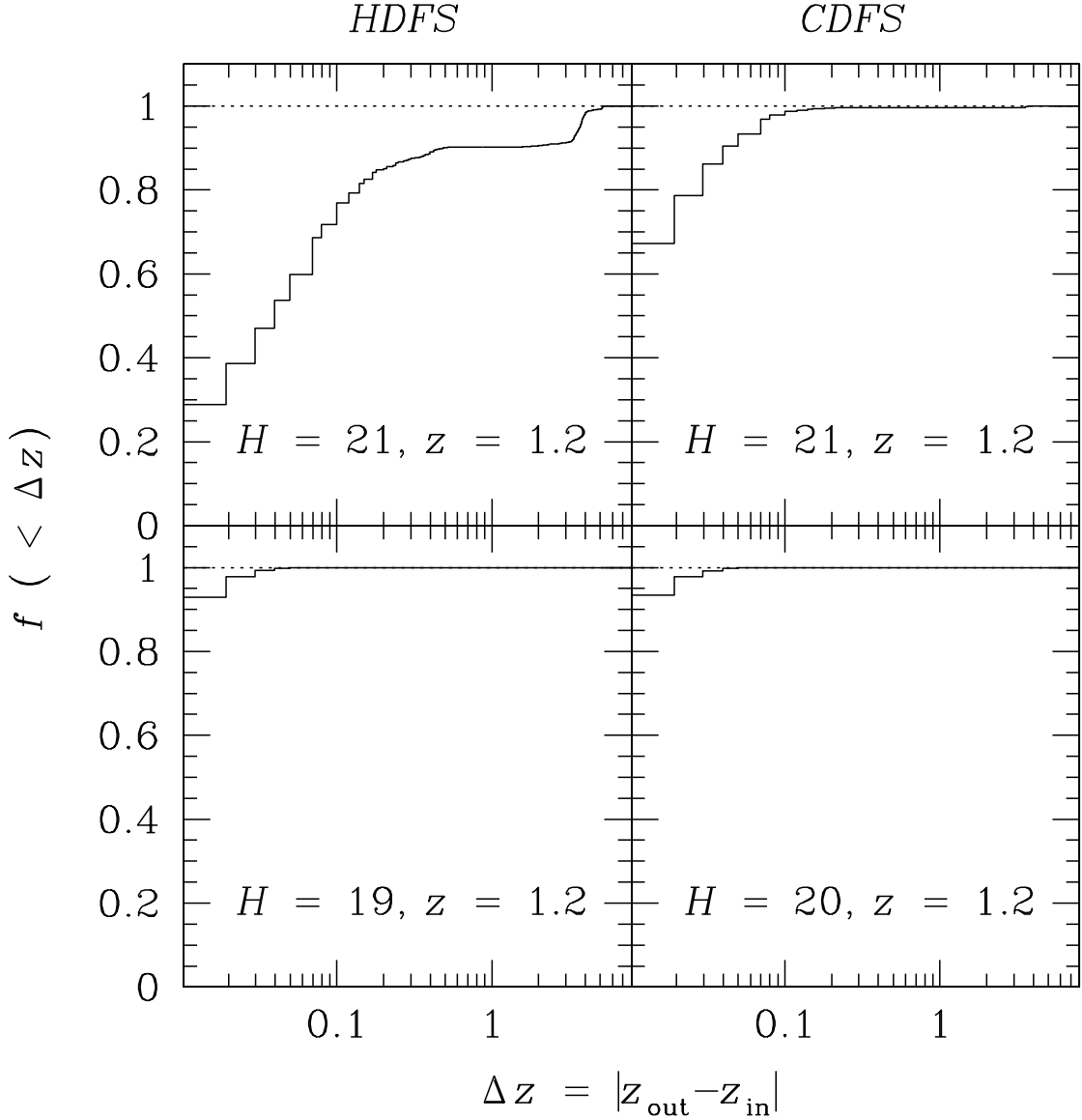}
\caption{Cumulative fractions of the simulated photometric redshifts versus
absolute redshift residuals for the four $z=1.2$ cases presented in Figure 5.
The left panels show the simulation results for the HDFS galaxies and the right
panels for the CDFS galaxies.  At $z=1.2$, it is clear that photometric 
redshifts are accurate to within $\Delta z=0.1$ for the CDFS
galaxies of $H\leq 21$, and that they become very uncertain for the HDFS 
galaxies at $H\sim 21$ because of their relatively larger photometric 
uncertainties.  Specifically, there is $\approx$ 30\% chance that the 
photometric redshifts of these galaxies are off by more than $\Delta z=0.1$.}
\end{figure}

\newpage

\begin{figure}
\epsscale{0.95}
\plotone{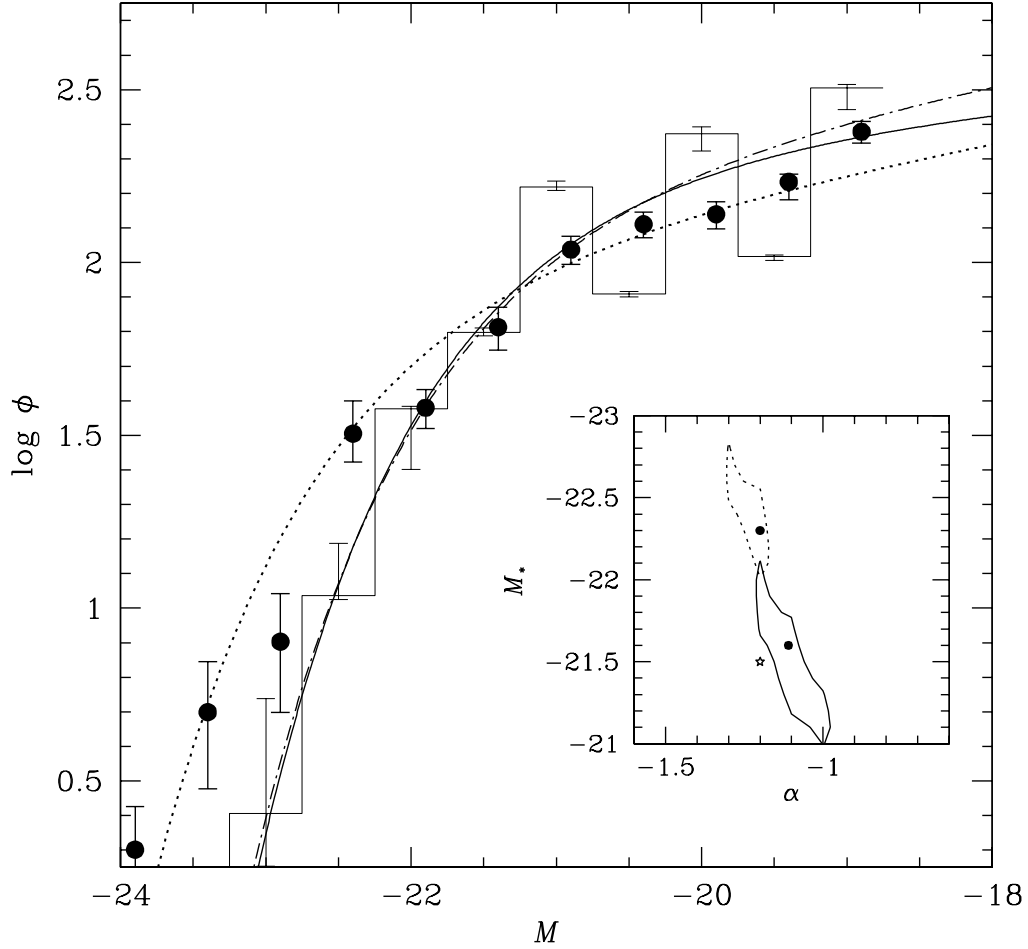}
\caption{Simulation results showing systematic uncertainties in the luminosity 
function calculations due to the relatively large redshift uncertainties.  The
dash-dotted curve indicates the input (intrinsic) galaxy luminosity function 
with the selected Schechter-function parameters, $M_*$ and $\alpha$, indicated
by the star in the inset.  The solid circles represent measurements using the 
``observed'' redshift catalog based on the $1/V_{\rm max}$ approach.  Error 
bars are the associated $1\,\sigma$ uncertainties estimated using a bootstrap 
re-sampling technique that takes into account both the sampling and redshift 
uncertainties.  The dotted curves represent the best-fit Schechter luminosity 
function based on the STY approach without including the redshift error 
functions of individual galaxies in the likelihood analysis.  The best-fit 
$M_*$ and $\alpha$ are indicated by the upper solid point in the inset with the
dotted contour showing the 99\% uncertainties.  The best-fit Schechter 
luminosity function based on our modified maximum likelihood analysis is shown
as the solid curve.  The best-fit $M_*$ and $\alpha$ are indicated by the lower
solid point in the inset with the solid contour showing the 99\% uncertainties.
The step function shows the results of the modified SWML method with the 
vertical bars indicating the associated $1\,\sigma$, one-parameter 
uncertainties.}
\end{figure}

\newpage

\begin{figure}
\epsscale{0.95}
\plotone{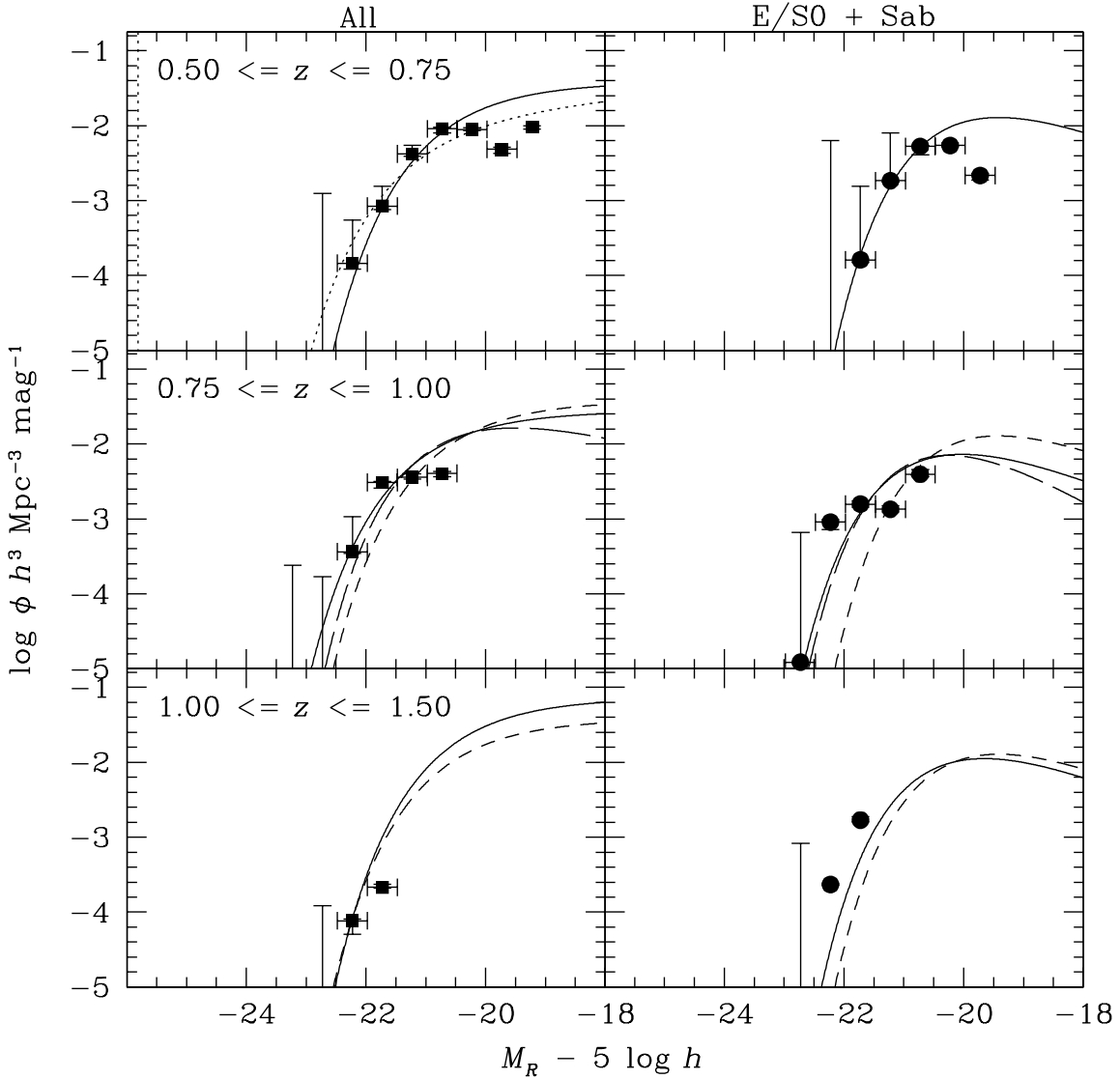}
\caption{The rest-frame $R$-band galaxy luminosity functions calculated using 
the modified STY (curves) and SWML (solid points) approaches.  The measurements
are for the total $H$-band selected galaxy sample (left panels) and for 
galaxies that have a best-fit spectral template of either an E/S0 or an Sab 
galaxy in the photometric redshift likelihood analysis (right panels).  The 
results shown in the top panels were obtained based on galaxies detected in the
HDFS regions only, excluding the CDFS galaxies at $z\leq 0.75$ for which the 
photometric redshifts become unreliable due to the lack of $U$ and $B$ 
photometry.  The horizontal bars indicate the bin size in absolute magnitude 
$\Delta M_R$.  The vertical bars associated with the solid points represent the
$1\,\sigma$, one-parameter uncertainties in the SWML analysis.  The solid and 
long-dashed curves represent, respectively, the best-fit Schechter luminosity 
functions with and without $\alpha$ fixed at the fiducial value at $0.5 \leq z 
\leq 0.75$.  The fiducial luminosity function at $0.5 \leq z \leq 0.75$ for 
each sub-sample is shown as the short-dashed curve in the middle and bottom 
panels for comparison.  For comparison, we also include in the top-left panel
the $r*$ luminosity function derived for galaxies at $z<0.2$ from the SDSS 
(Blanton \etal\ 2001; dotted curve).}
\end{figure}

\newpage

\begin{figure}
\plotone{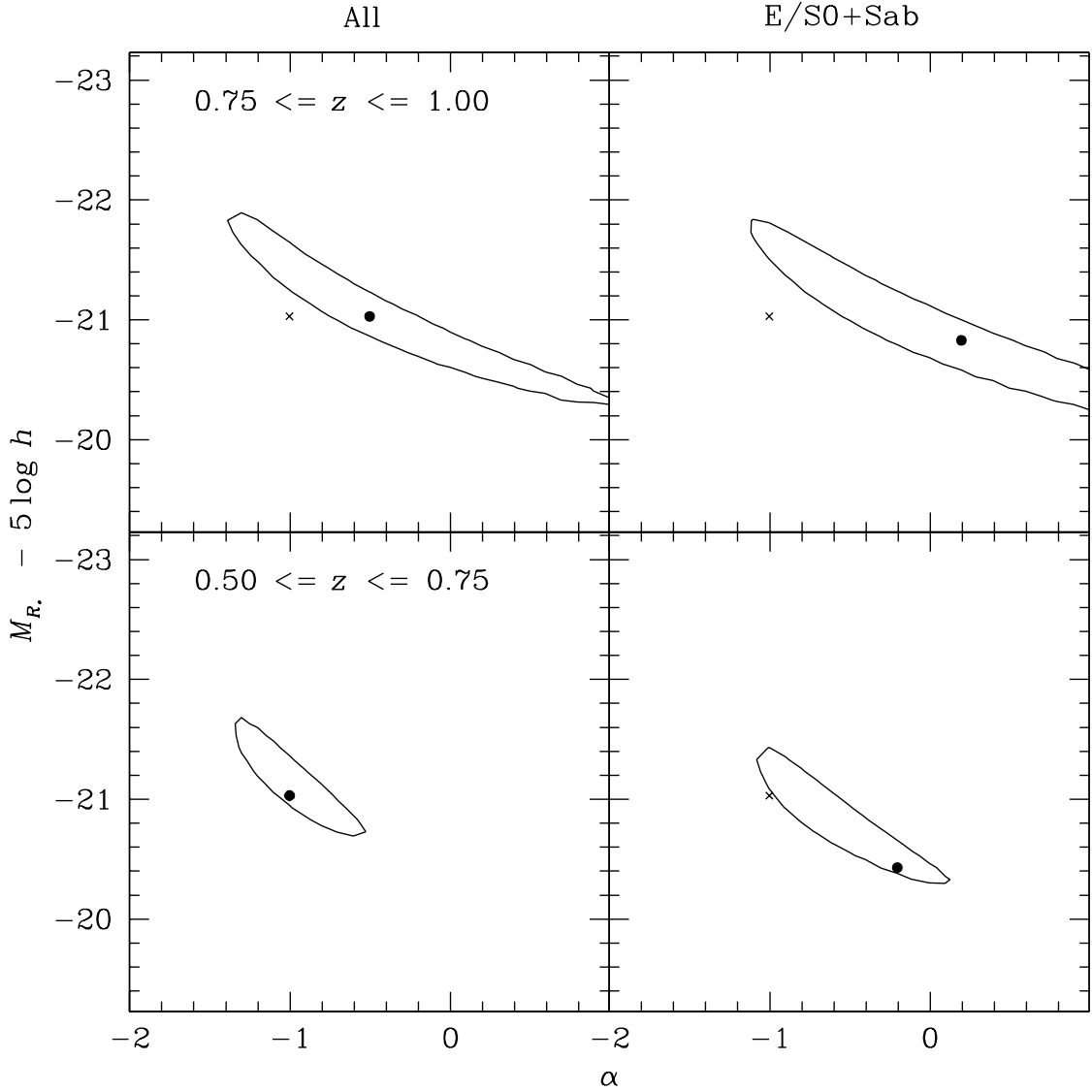}
\caption{The 99\% error contours of the best-fit $M_{R_*}$ and $\alpha$ in the
modified STY approach for the entire $H$-band selected galaxy sample (left 
panels) and the $H$-band selected early-type galaxies (right panels) at $0.5 
\leq z \leq 0.75$ (bottom panels) and $0.75 \leq z \leq 1.0$ (top panels).  The
cross in each panel indicates the best-fit $M_{R_*}$ and $\alpha$ for the total
$H$-band sample at $0.5 \leq z \leq 0.75$.}
\end{figure}

\newpage

\begin{figure}
\plotone{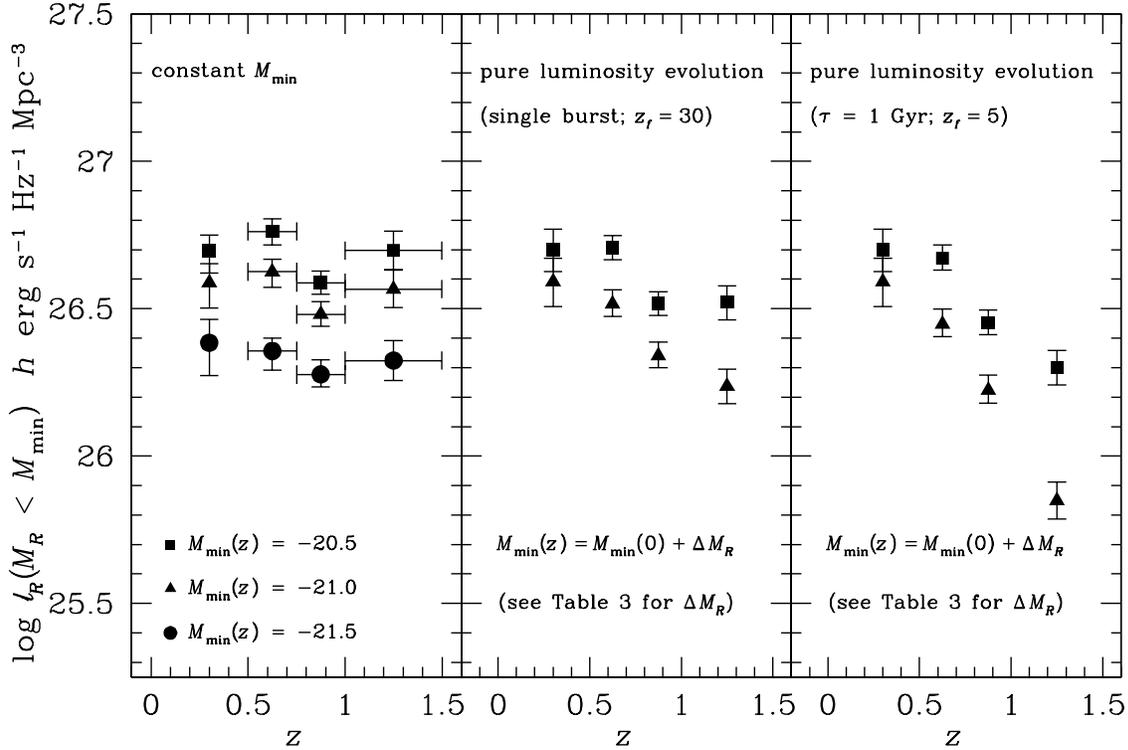}
\caption{Redshift evolution of the rest-frame co-moving $R$-band luminosity 
density $\ell_R$ for color-selected early-type galaxies brighter than $M_{\rm 
min}$.  We show in the left panel the results calculated by adopting a constant
$M_{\rm min}$ versus redshift for three different threshold values: $M_{\rm 
min}=-20.5$ in squares, $-21.0$ in triangles, and $-21.5$ in circles.  The 
$z=0.3$ values were derived based on the galaxy luminosity function determined
by Lin \etal\ (1999) from the CNOC2 early-type sample.  We also calculated 
$\ell_R$ by adopting an evolving $M_{\rm min}$ in order to remove the effect of
pure luminosity evolution of stars.  The middle panel shows the results for 
$M_{\rm min}$ estimated under a single burst scenario for a galaxy formed at 
$z_f\sim 30$.  The right panel shows the results for $M_{\rm min}$ estimated 
under a 1-Gyr exponentially declining star formation rate scenario for a galaxy
formed at $z_f\sim 5$.  The predicted brightening with redshift under different
scenarios is listed in Table 3.  The results in the two right panels have been
scaled to have consistent $M_{\rm min}$ at $z=0.3$ as those in the left panel.}
\end{figure}

\newpage

\begin{figure}
\plotone{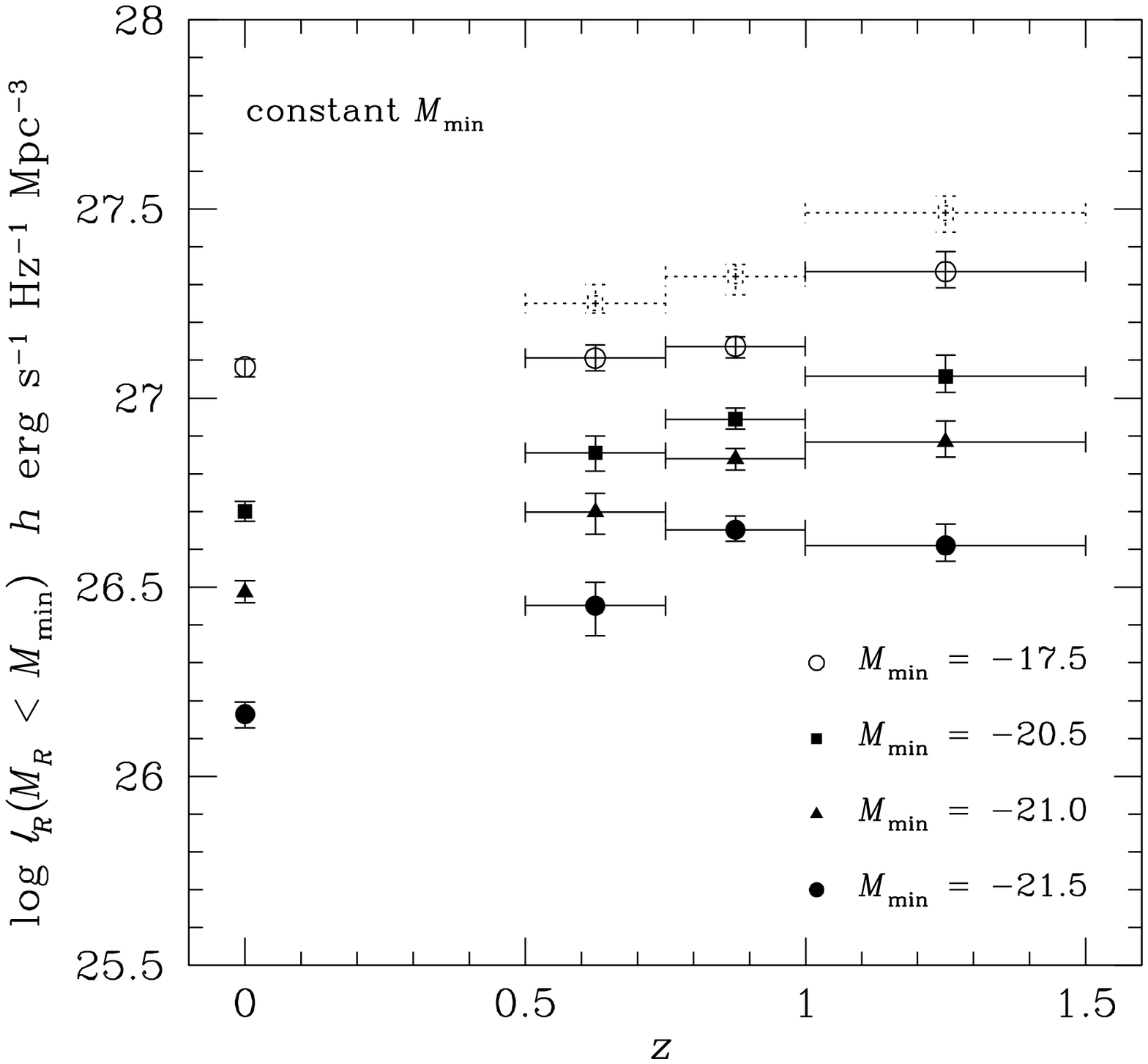}
\caption{Redshift evolution of the rest-frame co-moving $R$-band luminosity 
density for all $H$-band selected galaxies brighter than $M_{\rm min}$. 
The results were calculated by adopting a constant $M_{\rm min}$ versus 
redshift for three different threshold values: $M_{\rm min}=-20.5$ in squares, 
$-21.0$ in triangles, and $-21.5$ in circles.  The $z=0$ values were derived 
based on the galaxy luminosity function determined by Blanton \etal\ (2001) 
from the SDSS sample.  It shows that the luminosity density decreases from $z
\sim 1$ to $z\sim 0$ by approximately $\Delta\log l\,/\Delta\log (1+z) = 1.0 
\pm 0.2$ at 6800\,\AA.  To examine how sensitive this evolutionary slope is to
galaxies at the faint end, we extrapolated the calculation to $M_{\rm min} =
-17.5$ by adopting the best-fit $\alpha=-1.2$ of the SDSS luminosity function 
at $z=0$ and $\alpha=-1$ ($-1.5$) for the LCIR luminosity functions at higher 
redshifts.  The results are shown in open (dotted) circles and we find $\Delta
\log l\,/\Delta\log (1+z)\approx 0.6$--1.0.}
\end{figure}


\newpage

\begin{references}

\reference{} Bertin, E. \& Arnouts, S. 1996, A\&AS, 117, 393

\reference{} Binggeli, B., Sandage, A., \& Tammann, G. A. 1988, ARAA, 26, 509

\reference{} Blanton, M. R. \etal\ 2001, AJ, 121, 2358

\reference{} Brinchmann, J., Abraham, R., Schade, D., Tresse, L., Ellis, R. S.,
	     Lilly, S., Le Fevre, O., Glazebrook, K., Hammer, F., Colless, M.,
	     Crampton, D., \& Broadhurst, T. 1998, ApJ, 499, 112

\reference{} Bromley, B. C., Press, W. H., Lin, H., \& Kirshner, R. P., ApJ, 
	     505, 25

\reference{} Charlot, S. 1996, in The Universe at High-z, Large Scale 
	     Structure, and the Cosmic Microwave Background, ed. E. 
	     Martinez-Gonzalez \& J. L. Sanz (Heidelberg: Springer), p. 53

\reference{} Chen, H.-W., McCarthy, P. J., \& Marzke, R. O. \etal\ 2002, ApJ, 
	     570, 54

\reference{} Cohen, J. G., Hogg, D. W., Blandford, R., Cowie, L. L., Hu, E.,
	     Songaila, A., Shopbell, P., \& Richberg, K. 2000, ApJ, 538, 29

\reference{} Cohen, J. G. 2002, ApJ, 567, 672


\reference{} Connolly, A. J., Szalay, A. S., Dickinson, M., Subbarao, M. U.,
             \& Brunner, R. J. 1997, ApJ, 486, L11

\reference{} Connolly, A. J., Csabai, I., Szalay, A. S., Koo, D. C., Kron, R.
	     G., \& Munn, J. A. 1995, AJ, 110, 2655

\reference{} Cowie, L. L., Songaila, A., \& Barger, A, J. 1999, AJ, 118, 603

\reference{} Cowie, L. L., Songaila, A., Hu, E. M., \& Cohen, J. G. 1996, AJ,
                112, 839

\reference{} Cross, N., Driver, S. P., Couch, W., Baugh, C. M., Bland-Hawthorn,
	     J., Bridges, T., Cannon, Rs., Cole, S., Colless, M., Collins, C.,
	     Dalton, G., Deeley, K., de Propris, R., Efstathiou, G., Ellis, 
	     R. S., Frenk, C. S., Glazebrook, K., Jackson, C., Lahav, O., 
	     Lewis, I., Lumsden, S., Maddox, S., Madgwick, D., Moody, S., 
	     Norberg, P., Peacock, J. A., Peterson, B. A., Price, I., Seaborne,
	     M., Sutherland, W., Tadros, H., \& Taylor, K. 2001, MNRAS, 324, 
	     825

\reference{} Cristiani, S., Appenzeller, I., Arnouts, S., Nonino, M., 
	     Arag\'on-Salamanca, A., Benoist, C., da Costa, L., Dennefeld, M.,
	     Rengelink, R., Renzini, A., Szeifert, T., \& White, S. 2000, A\&A,
	     359, 489

\reference{} Daddi, E., Cimmatti, A., Pozzetti, L., Hoekstra, H., R\"ottgering,
             H. J. A., Renzini, A., Zamorani, G., \& Mannucci, F. 2000, A\&A,
             361, 535

\reference{} de Lapparent, V., Geller, M. J., \& Huchra, J. P. 1989, ApJ, 343,
	     1

\reference{} Dennefeld, M. 2002, http://www.iap.fr/hst/tmrresults.html

\reference{} Efstathiou, G., Ellis, R. S., \& Peterson, B. A. 1988, MNRAS, 232,
	     431

\reference{} Eggen, O. J., Lynden-Bell, D., \& Sandage, A. R. 1962, ApJ, 136,
             748

\reference{} Ellis, R. S., Colless, M., Broadhurst, T., Heyl, J., \&
	     Glazebrook, K. 1996, MNRAS, 280, 235

\reference{} Elston, R., Rieke, G. H., \& Rieke, M. J. 1988, ApJ, 331, L77

\reference{} Fan, X. \etal\ 2000, AJ, 119. 928

\reference{} Felten, J. E. 1976, ApJ, 207, 700

\reference{} Fern\'andez-Soto, A., Lanzetta, K. M., \& Yahil, A. 1999, ApJ,
             513, 34

\reference{} Fern\'andez-Soto, A., Lanzetta, K. M., Chen, H.-W., Pascarelle, S.
             M., \& Yahata, N. 2001, ApJS, 135, 41

\reference{} Fern\'andez-Soto, A., Lanzetta, K. M., Chen, H.-W., Levine, B., \&
	     Yahata, N. 2002, MNRAS, 330, 889


\reference{} Firth, A. E., Somerville, R., \& McMahon, R. G. \etal\ 2002, 
	     MNRAS, 332, 617

\reference{} Francis, P. J., Hewett, P. C., Foltz, C. B., Chaffee, F. H.,
             Weymann, R. J., \& Morris, S. L. 1991, ApJ, 373, 465

\reference{} Fried, J. W., von Kuhlmann, B., Meisenheimer, K., Rix, H.-W.,
	     Wolf, C., Hippelein, H. H., K\"ummel, M., Phleps, S., R\"oser, H.
	     J., Thierring, I., \& Maier, C. 2001, A\&A, 367, 788

\reference{} Gardner, J. P., Sharples, R. M., Frenk, C. S., \& Carrasco, B. E.
             1997, ApJ, 480, L99 

\reference{} Glazebrook, C. 1998,  http://www.aao.gov.au/hdfs/Redshifts

\reference{} Groth, E. \etal, 1994, BAAS, 185, 5309

\reference{} Gwyn, S. D. J. \& Hartwick, F. D. A. 1996, ApJ, 468, L77

\reference{} Heyl, J., Colless, M., Ellis, R. S., \& Broadhurst, T. 1997, 
	     MNRAS, 285, 613


\reference{} Hu, E. M. \& Ridgway, S. E. 1994, AJ, 107, 1303

\reference{} Im, M., Griffiths, R. E., Naim, A., Ratnatunga, K. U., Roche, 
	     N., Green, R. F., \& Sarajedini, V. L. 1999, ApJ, 510, 82

\reference{} Kauffmann, G., Charlot, S., \& White, S. D. M. 1996, MNRAS, 283,
             117


\reference{} Lanzetta, K. M., Yahil, A., \& Fern\'{a}ndez-Soto, A. 1996,
                Nature, 381, 759

\reference{} Lanzetta, K. M., Fern\'andez-Soto, A., \& Yahil, A. 1998,
             in ``The Hubble Deep Field, Proceedings of the Space Telescope
             Science Institute 1997 May Symposium,'' ed.\ M. Livio, S. M. Fall,
             \& P. Madau (Cambridge:  Cambridge University Press), P. 143

\reference{} Lanzetta, K. M., Chen, H.-W., Fern\'andez-Soto, A., Pascarelle, 
	     S., Puetter, R., Yahata, N., \& Yahil, A. 1999, in ``Photometric
             Redshifts and High Redshift Galaxies'', eds. R. Weymann, L.
             Storrie-Lombardi, M. Sawicki, \& R. Brunner, P. 223

\reference{} Leggett, S. K., Allard, F., Dahn, C., Hauschildt, P. H., Kerr, T.
             H., \& Rayner, J. 2000, ApJ, 535, 965

\reference{} Lilly, S. J., Le F\`evre, O., Hammer, F., \& Crampton, D. 1996,
                ApJ, 461, 534

\reference{} Lilly, S. J., Tresse, L., Hammer, F., Crampton, D., \& Le F\`evre,
	     O. 1995, ApJ, 455, 108

\reference{} Lin, H., Yee, H. K. C., Carlberg, R. G., Morris, S. L., Sawicki,
             M., Patton, D. R., Wirth, G., \& Shepherd, C. W. 1999, ApJ, 518,
             533

\reference{} Madgwick, D.S., Lahav, O., Baldry, I.K., Baugh, C.M., 
	     Bland-Hawthorn, J., Bridges, T., Cannon, R., Cole, S., Colless, 
	     M., Collins, C., Couch, W., Dalton, G., De Propris, R., Driver, 
	     S., Efstathious, G., Ellis, R.S., Frenk, C.S., Glazebrook, K., 
	     Jackson, C., Lewis, I., Lumsden, S., Maddox, S., Norberg, P., 
   	     Peacock, J.A., Peterson, B.A., Sutherland, W., \& Taylor, K. 2001,
	     astro-ph/0107197

\reference{} Martini, P. 2001, AJ, 121, 2301

\reference{} Marzke, R. O., Huchra, J. P., \& Geller, M. J. 1994, ApJ, 428, 43

\reference{} Marzke, R. O. \& da Costa, L. N. 1997, AJ, 113, 185

\reference{} Marzke, R. O., da Costa, L. N., Pellegrini, P. S., Willmer, C. N.
	     A., \& Geller, M. J. 1998, ApJ, 503, 617

\reference{} Marzke, R. O., McCarthy, P. J., \& Persson, S. E. \etal\ 1999,
             in "Photometric Redshifts and High Redshift Galaxies", eds. R.
             Weymann, L. Storrie-Lombardi, M. Sawicki, \& R. Brunner. A.S.P.
             Conf. Series vol.\ 191, p. 148

\reference{} McCarthy, P. J., Persson, S. E. \& West, S. C. 1992, ApJ, 386, 52

\reference{} McCarthy, P. J., Carlberg, R. G., Chen, H.-W., \& Marzke, R. O.
             \etal\ 2001, ApJ, 560, L131

\reference{} Menanteau, F., Ellis, R. S., Abraham, R. G., Barger, A. J., \&
             Cowie, L. L. 1999, MNRAS, 309, 208

\reference{} Mobasher, B., Rowan-Robinson, M., Georgakakis, A., \& Eaton, N.
	     1996, MNRAS, 282, L7

\reference{} Mobasher, B., Ellis, R. S., \& Sharples, R. M. 1986, MNRAS, 223,
             11 

\reference{} Oppenheimer, B. R., Kulkarni, S. R., Matthews, K., \& van
             Kerkwijk, M. H. 1998, ApJ, 502, 932

\reference{} Palunas, P., Collins, N. R., Gardner, J. P., Hill, R. S., 
		Malumuth, E. M., Smette, A., Teplitz, H. I., Williger, G. M., 
		\& Woodgate, B. E. 2000, ApJ, 541, 9

\reference{} Pickles, A. J. 1998, PASP, 110, 863

\reference{} Poli, F., Menci, N., Giallongo, E., Fontana, A., Cristiani, S.,
	     \& d'Odorico, S. 2001, ApJ, 551, L45

\reference{} Rudnick, G., Franx, M., Rix, H.-W., Moorwood, A., Kuijken, K.,
	     van Starkenburg, L., van der Werf, P., R\"ottgering, H., van 
	     Dokkum, P., \& Labbé, I. 2001, AJ, 122, 2205

\reference{} Sandage, A., Tammann, G. A., \& Yahil, A. 1979, ApJ, 232, 352

\reference{} Sawicki, M. J., Lin, H., \& Yee, H. K. C. 1997, AJ, 113, 1

\reference{} Schmidt, M. 1968, ApJ, 151, 393

\reference{} Shleqey, A. E., Steidel, C. C., Adelberger, K. L., Dickinson, M.,
		Giavalisco, M., \& Pettini, M. 2001, ApJ, 562, 95

\reference{} Subbarao, M. U., Connolly, A. J., Szalay, A. S., \& Koo, D. C.
		1996, AJ, 112, 929

\reference{} Totani, T. \& Yoshii, J. 1997, ApJ, 501, L177

\reference{} Tresse, L., Dennefeld, M., Petitjean, P., Cristiani, S., \& White,
		S. 1999, A\&A, 346, L21

\reference{} White, S. D. M. \& Rees, M. J. 1978, MNRAS, 183, 341

\reference{} Yahata, N., Lanzetta, K.M., Chen, H.-W., Fern\'andez-Soto, A.,
             Pascarelle, S.M., Puetter, R.C., \& Yahil, A. 2000, ApJ, 538, 493

\reference{} Zheng, W., Kriss, G. A., Telfer, R. C., Grimes, J. P., \&
             Davidsen, A. F. 1997, ApJ, 475, 469


\end{references}
\end{document}